\documentclass[a4paper,12pt]{article}
\usepackage[utf8]{inputenc}
\usepackage[english]{babel}
\date{} 
\usepackage{ragged2e}
\usepackage{multirow}
\usepackage{rotating} 
\usepackage{amsmath}
\usepackage{graphicx}
\usepackage{diagbox}
\usepackage{makecell} 
\usepackage{subfigure}
\usepackage{epsfig}
\usepackage{ragged2e}
\usepackage{booktabs}
\usepackage{soul}
\usepackage{xcolor}
\usepackage{amssymb}
\usepackage{graphicx}
\usepackage{enumerate}
\usepackage{enumitem}
\usepackage{array}
\usepackage{lscape}
\usepackage{authblk}
\usepackage{multicol}
\usepackage{adjustbox}
\usepackage{csquotes}
\usepackage{float}
\usepackage{afterpage}
\usepackage{graphicx,subcaption} 
\usepackage{sectsty}
\usepackage{array}
\usepackage{booktabs}
\usepackage{multirow}
\usepackage{tabularx}
\usepackage{tikz} 
\usetikzlibrary{positioning,arrows.meta}
\usepackage{array}
\usepackage{longtable}
\usepackage{listings} 
\usepackage{xcolor}
\lstdefinestyle{Rcompact}{
	language=R,
	basicstyle=\scriptsize\ttfamily,
	stepnumber=1,
	numbersep=4pt,
	breaklines=true,
	breakatwhitespace=false,
	showstringspaces=false,
	keepspaces=true,
	columns=fullflexible,
	frame=none,
	xleftmargin=15pt,
	aboveskip=3pt,
	belowskip=3pt,
	lineskip=-1pt
}

\numberwithin{equation}{section}
\numberwithin{table}{section}
\numberwithin{figure}{section}
\usepackage{booktabs}\usepackage{dcolumn}\usepackage{caption}
\usepackage{threeparttable}\usepackage{booktabs}
\usepackage{rotating}\usepackage{booktabs}\usepackage{ulem}
\usepackage{indentfirst}\usepackage{array}\usepackage{times}\usepackage{color}\usepackage{psfrag}\usepackage{eso-pic}
\usepackage{type1cm}
\usepackage[margin=2.2cm]{geometry}
\date{}
\title{ \bfseries\Large Analysis of Nonnegative Observations using Gamma Model with 2 Factors (ANOGaM-2):
Theory, Method and Applications with Real-life Data  (including R code)}
\author[1]{Buu-Chau Truong}
\author[1]{Nuong Thi Thuy Tran}
\author[1,2*]{Nabendu Pal}
\affil[1]{\small Faculty of Mathematics and Statistics, Ton Duc Thang University, Ho Chi Minh City, Vietnam}
\affil[2]{\small Department of Mathematics, University of Louisiana at Lafayette, USA}
\affil[*]{\normalsize  Corresponding author: \it{nabendu.pal@tdtu.edu.vn}; \it{\normalsize nabendu.pal@louisiana.edu}}
\date{}
\begin{document}
\maketitle \setcounter{page}{1}
\vspace{-1.35cm}
\begin{abstract}
A two-factor analysis of variance (ANOVA) is widely used in design of experiments when experimental units are subjected to two factors (i.e., potential sources of variations). However, such an analysis, which uses the $F$-tests, is dependent on three critical assumptions, - (i) all the main and interaction effects as well as the unexplained error term are additive to represent the response variable; (ii) the unobservable errors are all independent and follow a normal distribution; and (iii) the error variances, though unknown, are all equal (i.e., the errors are homoscedastic). In many engineering and biological studies, where the observations are nonnegative to begin with, it is often found that one or more of the above assumptions is/are not tenable. Further, the observations tend to exhibit positively skewed distributions as seen from the sample histograms. In such situations, the standard operating procedure (SOP) of the two-factor ANOVA calls for a suitable (Box-Cox type) transformation, so that the transformed observations can follow the aforementioned model assumptions. There are two practical difficulties faced by the researchers with the transformed observations: (a) the transformed observations lose their relevance to the original problem, and the resultant unit(s) of the transformed observations can be meaningless; and (b) it becomes a subjective call to come up with the most appropriate transformation of the data, i.e., one transformation can make the data adhere to one assumption while another transformation can make the data follow another assumption closely. Faced with such a dilemma (where the observations tend to be positively skewed) we offer a paradigm shift where the nonnegative observations, influenced by two factors, are modeled by gamma distributions with unknown shape and scale parameters which may depend on the corresponding factor levels. We then proceed with testing the main effects (whether the main effects of a factor are all equal or not) and interaction effects (whether the interactions exist or not). To test a null hypothesis against a suitable alternative, we first derive the likelihood ratio test (LRT) statistic, and study its performance based on its asymptotic Chi-square distribution. But since the asymptotic LRT (henceforth called `ALRT') may not work very well for small to moderate sample sizes, we then propose a parametric bootstrap (PB) test based on the LRT statistic which does not use the Chi-square distribution, rather finds its critical value automatically through simulation. The PB test using the LRT statistic (henceforth called `PBLRT') appears to work remarkably well in terms of maintaining the nominal level as seen from our comprehensive simulation study. Further, we present some real-life datasets to buttress the applicability of our proposed PBLRT over the classical ALRT, and to demonstrate how the inferences may differ from the ones based on traditional normality and homoscedasticity based ANOVA.
\end{abstract}
\vspace{-0.1cm}
\vspace{0.3cm}

\linespread{1.2}
\fontsize{12pt}{16pt}\selectfont

{\section{Introduction}
\subsection{Preliminaries}	
The two-factor Analysis of Variance (ANOVA) is a widely used parametric method to study the effect of two factors (i.e., potential sources of variation) on a response variable. Suppose the experimental response variable is exposed to two factors, say $A$ and $B$, with levels $A_1, A_2,...,A_a$ and $B_1, B_2,...,B_b$ respectively, then the standard theory assumes the following additive linear model
\begin{equation}
X_{ijk}=\mu+\tau_i+\beta_j+\gamma_{ij}+\varepsilon_{ijk},
\end{equation}
where $X_{ijk}$ is the $k^{th}$ replication of the response variable subject to $A_i$ and $B_j;1\le i\le a; 1\le j \le b;$ and $1\le k \le n_{ij}.$ In (1.1), $\mu$ is the general effect, $\tau_i$ is the main effect of $A_i$, $\beta_j$  is the main effect of $B_j$, and $\gamma_{ij}$ is the interaction effect due to the combination $(A_i, B_j)$. The standard ANOVA technique further assumes that the unobservable error terms $\varepsilon_{ijk}\textrm{'}s$ are independent and identically distributed random variables following $N(0,\sigma^2)$. The equality of variances of $\varepsilon_{ijk}$ over all $(i,j)$ is an important assumption known as `homoscedasticity'.

Under the above assumptions, the standard two-way ANOVA theory calls for decomposing the `total sum of squares' (or $SS_{Total}$), defined as 
$SS_{Total}=\sum_{i=1}^a \sum_{j=1}^b \sum_{k=1}^{n_{ij}}(X_{ijk}-\bar X_{\cdots})^2,$ with $\bar X_{\cdots}$ = the grand mean of all observations, as
\begin{equation}
SS_{Total}=SS_A+SS_B+SS_{AB}+SS_{Error},
\end{equation}
where the terms in (1.2) are known as the sum of squares due to Factor-$A$, due to Factor-$B$, due to the interaction between $A$ and $B,$ and the sum of squares due to error, respectively. Further, $SS_A, SS_B$ and $SS_{AB}$ can be standardized suitably by using $SS_{Error}$ in order to test the significance of the main effects as well as the interaction effect.

The standard text books as well as popular softwares have discussions or inbuilt mechanisms to check the model assumptions. Slight departures from the assumptions are not considered serious, and the null distributions of the relevant test statistics do follow the required $F$-distributions fairly closely. But what happens when there are severe departures from the model assumptions? Before we talk about what to do when there are departures from the model assumptions, we have to keep in mind that detection of these departures are often inaccurate since such remedial tests themselves are often asymptotic in nature (i.e., works well only for large sample sizes) and not very effective for small sample sizes. When there are suspected departures from the model assumptions, one is expected to transform the data suitably, often through the `Box - Cox method' so that the transformed response variable can follow the model assumptions more closely. However, such a transformation can make the resultant response variable hard to interpret, and the associated transformed measuring unit may become meaningless.

Note that under the linear additive model (1.1), the observation $X_{ijk}\sim N(\mu+\tau_i+\beta_j+\gamma_{ij},\sigma^2)$, and due to over parametrization we further assume that
$\sum_{i=1}^a \tau_i= \sum_{j=1}^b \beta_j =\sum_{i=1}^a \gamma_{ij}=\sum_{j=1}^b \gamma_{ij}=0.$ But in many engineering and biological studies, the observations $X_{ijk}, (1\le k \le n_{ij})$ tend to be positively skewed. At the least, the normality of the observations $X_{ijk}$ can't be assured with a high degree of certainty for small sample sizes. In other words, the standard normality tests like the `Wilk - Shapiro test' or `Anderson - Darling test' do not have high power for small $n_{ij}$'s even when there is a significant departure from normality. Further, the assumptions of linear additivity (as in (1.1)) and  homoscedasticity may not be justified as well as verified with a high level of certainty for small sample sizes.
\subsection{A Few Motivating Examples}

To underscore the above deficiencies with the standard practice of a two-way ANOVA, we first present a few real-life datasets in the following.

\textbf{Example 1.1} Juskevich and Guyer (1990) reported the weight gains (= response variable) in rats over a period of eighty five days in six treatment conditions of recombinant bovine growth hormone (rbgh) as shown in Table 1.1. The observations were recorded according to gender (Factor-$A$, say) and the hormone (Factor-$B$, say) with $n_{ij}=5$.

\begin{table} [H]
	\caption{ Weight gains in rats according to gender and hormone.}
	\centering
		\begin{tabular}	{|*{7}{c|}} \hline
		\multirow{2}{*}{Gender (Factor - $A$)}
		&\multicolumn{6}{c|}{Growth hormone (Factor - $B$)}\\	
		\cline {2-7}
		&1&2&3&4&5&6\\ \hline
		\multirow{5}{*}{1} &274.99& 478.62&276.32 &391.45 &405.58 &264.89\\
		&289.67&399.14 &318.32&351.85&318.63&308.36 \\
		&346.40&512.03&310.48&299.05&291.58&338.21\\ &344.32 & 400.74 & 352.29&288.55&311.40&352.18 \\ 
		&364.63 &369.48 &377.58 &259.10&297.80&376.36\\ \hline
		
		\multirow{5}{*}{2} &153.96& 190.02&111.37 &120.05 &118.60 &149.38\\
		&153.92&253.14 &139.92&127.68&153.60&143.54 \\
		&181.88&176.91&166.74&145.49&130.52&124.30\\ &132.75 & 227.31 & 140.66&172.93&159.76&164.93 \\ 
		&117.50 &237.62 &141.31 &193.85 & 172.52&177.85\\ \hline
	\end{tabular}
\end{table}
	
\textbf{Example 1.2} The following Table 1.2 provides the `total color change score' of two brands, each with three formulations, of self-tanning products as reported by Jermann, Toumiat and Imjeld (2002) (with $n_{ij}=4$).
\begin{table}[H] \caption{Self-tanning products' total color change score.}
	\centering
	\begin{tabular}	{|*{4}{c|}} \hline
		\multirow{2}{*}{Brand (Factor - $A$)}
		&\multicolumn{3}{c|}{Formulation (Factor - $B$)}\\	
		\cline {2-4}
		&1&2&3\\ \hline
		\multirow{2}{*}{1} &16.79; 12.68; &10.23; 10.29;
		 &9.43; 9.45\\
		&12.47; 11.67; &8.97; 8.51;&8.86; 8.66\\
		 \hline
		\multirow{2}{*}{2} &32.85; 38.08;&25.06; 21.66; &25.89; 22.96\\
		&30.25; 28.41; &19.86; 18.62; & 24.55; 24.59\\ \hline
	\end{tabular}
	
\end{table}

\textbf{Example 1.3} The following Table 1.3 presents the data obtained by Azcel (1996) comparing the brightness of films produced by three different manufacturers and developed using three different processes (with $n_{ij}=5$).

\begin{table}[ht]\caption{Measurement on brightness of films.}
	\centering
		\begin{tabular}	{|*{4}{c|}} \hline
			\multirow{2}{*}{Manufacturer (Factor - $A$)}
			&\multicolumn{3}{c|}{Development Process (Factor - $B$)}\\	
			\cline {2-4}
			&1&2&3\\ \hline
		    Kodak&32; 34; 31; 30; 37; &26; 29; 27; 30; 31;
			&28; 28; 27; 30; 32\\
		
			\hline
			Fuji &43; 41; 44; 50; 47;&32; 38; 38; 40; 36; &32; 32; 36; 35; 34\\	 \hline
			Agfa &23; 24; 25; 21; 26;&27; 30; 25; 25; 27; &25; 27; 26; 22; 25\\	 \hline
		\end{tabular}
	
\end{table}

\textbf{Example 1.4} The following Table 1.4 (from Crampton(1947)) presents the observations from 60 guinea pigs that were randomly designed to receive one of the three of ascorbic acid supplement, and one of two delivery methods-orange juice (OJ) or vitamin C (VC). The dependent variable is the tooth length, and each treatment combination is assigned to $n_{ij}=10$ guinea pigs.

\begin{table}[ht]\caption{Raw tooth length measurements by supplement type and dose.}
	\centering
	\setlength{\tabcolsep}{5pt}
	\fontsize{10pt}{12pt}
	\begin{tabular}{|c|c|c|c|}
		\hline
		\multirow{2}{*}{\begin{tabular}{c}
				Delivery method \\
				(Factor - $A$)
		\end{tabular}}
		&
		\multicolumn{3}{c|}{Supplement (Factor - $B$)}
		\\ \cline{2-4}
		& 0.5 mg/day & 1.0 mg/day & 2.0 mg/day \\
		\hline
		OJ &
		\begin{tabular}{l}
		13.6; 14.5; 10.0; 8.2; \\
		 9.4; 16.5; 9.7; 19.7;  \\ 
		  23.3; 23.6;
		
		\end{tabular}
		&
		\begin{tabular}{l}
			22.4; 24.5; 24.8; 30.9;\\
			 29.5; 21.5; 23.3; 23.6;\\ 
			 26.4; 20.0;
		\end{tabular}
		&
		\begin{tabular}{l}
			26.4; 32.5; 26.7; 21.5;\\
			 23.3; 29.5; 25.5; 26.4;\\ 
			 22.4; 24.8;
		\end{tabular}
		\\
		\hline
		
		VC &
		\begin{tabular}{l}
			4.2; 11.5; 7.3; 5.8; \\
			6.4; 10.0; 11.2; 11.2; \\
			 5.2; 7.0;
		\end{tabular}
		&
		\begin{tabular}{l}
			16.5; 16.5; 15.2; 17.3; \\
			22.5; 17.3; 13.6; 14.5; \\
			18.8; 15.5;
		\end{tabular}
		&
		\begin{tabular}{l}
			23.6; 18.5; 33.9; 25.5; \\
		26.4;32.5; 26.7; 21.5;\\ 
		23.3; 29.5;
		\end{tabular}
		\\
		\hline
		
	\end{tabular}
\end{table}

\textbf{Example 1.5} The following Table 1.5 presents the enzyme activity of mannose-6-phosphate isomerase (MPI) and MPI genotypes in the amphipod crustacean Platorchestia platensis (see McDonald (2014)).

\begin{table}[H]	\caption{Response values for each combination of genotype and gender.}
	\centering
	\setlength{\tabcolsep}{5pt}
	
	\begin{tabular}{|c|p{3.8cm}|p{3.8cm}|p{3.8cm}|}
		\hline
			\multirow{2}{*}{\begin{tabular}{c}
				Gender\\
				(Factor - $A$)
		\end{tabular}}
		&
		\multicolumn{3}{c|}{Genotype (Factor - $B$)}
		\\ \cline{2-4}
		& \centering FF	& \centering FS
		& \centering SS\arraybackslash \\
		\hline
		
		Female
		&
		2.838; 4.216; 2.889; 4.198;
		&
		3.550; 4.556; 3.087; 1.943;
		&
		3.620; 3.079; 3.586; 2.669;
		\\
		\hline
		
		Male
		&
		1.884; 2.823; 4.939; 3.486;
		&
		2.396; 2.956; 3.105; 2.649;
		&
		2.801; 3.421; 4.275; 3.110;
		\\
		\hline
		
	\end{tabular}

\linespread{1.5}\end{table}

\textbf{Example 1.6} The following Table 1.6 presents the bonding strength (MPa) of four types of dental resin (A, B, C, and D) under two curing light sources (Halogen and LED). The dataset with $n_{ij}=10$ for each combination of resin and light source has been taken from Kim (2014).

\begin{table}[H]\caption{Bonding strength measurements: four different types of resin and two curing methods.}
	\centering
	\fontsize{10pt}{12pt}
	\renewcommand{\arraystretch}{1.4}
	\setlength{\tabcolsep}{5pt}
		\begin{tabular}{|c|c|c|c|c|}
			\hline
			
			\multirow{2}{*}{\begin{tabular}{c}
					Curing light\\
					(Factor - $A$)
			\end{tabular}}
			&
			\multicolumn{4}{c|}{Resin type (Factor - $B$)}
			\\ \cline{2-5}
			
			& A & B & C & D \\
			\hline
			
			Halogen
			&
			\begin{tabular}{l}
				14.5; 15.2; 17.4;\\
				17.5; 19.2; 19.7;\\
				20.1; 21.3; 23.5;\\
				9.3
			\end{tabular}
			&
			\begin{tabular}{l}
				11.8; 13.3; 19.2;\\
				21.3; 22.2; 23.0;\\
				24.5; 24.6; 27.1;\\
				12.0
			\end{tabular}
			&
			\begin{tabular}{l}
				14.5; 15.0; 18.6;\\
				19.6; 21.0; 21.6;\\
				25.5; 25.9; 30.7;\\
				33.0
			\end{tabular}
			&
			\begin{tabular}{l}
				35.5; 35.7; 36.3;\\
				37.3; 39.9; 40.9;\\
				41.0; 44.5; 44.7;\\
				47.2
			\end{tabular}
			\\
			\hline
			
			LED
			&
			\begin{tabular}{l}
				27.1; 11.6; 12.2;\\
				15.9; 17.0; 17.2;\\
				18.4; 19.8; 23.4;\\
				28.0
			\end{tabular}
			&
			\begin{tabular}{l}
				27.8; 12.8; 16.3;\\
				19.8; 22.4; 23.6;\\
				25.3; 27.9; 34.6;\\
				35.2
			\end{tabular}
			&
			\begin{tabular}{l}
				16.5; 22.7; 24.2;\\
				26.2; 28.4; 28.5;\\
				30.7; 32.2; 33.8;\\
				34.5
			\end{tabular}
			&
			\begin{tabular}{l}
				17.3; 19.2; 19.5;\\
				20.5; 20.7; 22.2;\\
				25.8; 29.0; 29.2;\\
				35.1
			\end{tabular}
			\\
			\hline
			
		\end{tabular}
	\label{tab:bonding}
\end{table}

\textbf{Example 1.7} The following Table 1.7 shows the emotional maturity  scores of 27 young adult males cross-classified by age and the extent to which they use marijuana (see Daniel (1987), page-325).
\begin{table}[H]
	\centering
	\caption{Marijuana usage by age group.}
	\label{tab:marijuana_age}
	\renewcommand{\arraystretch}{1.3}
	
	\begin{tabular}{|c|c|c|c|}
		\hline
		\multirow{2}{*}{\begin{tabular}{c}
				Age \\
				(Factor - $A$)
		\end{tabular}}
		& \multicolumn{3}{c|}{Marijuana Usage (Factor-$B$)} \\
		\cline{2-4}
		& Never & Occasionally & Daily \\
		\hline
		
		15--19
		& 25; 28; 22
		& 18; 23; 19
		& 17; 24; 19 \\
		
		\hline
		
		20--24
		& 28; 32; 30
		& 16; 24; 20
		& 18; 22; 20 \\
		
		\hline
		
		25--29
		& 25; 35; 30
		& 14; 16; 15
		& 10; 8; 12 \\
		
		\hline
	\end{tabular}
\end{table}

\textbf{Example 1.8} The nutritive value of a certain edible fruit was measured in a total of 72 specimens consisting of six specimens of each of four varieties grown in each of three geographic regions with 
the data as given below in Table 1.8 (from Daniel (1987), page-327).
\begin{table}[H]
	\centering
	\caption{Nutritive values of four varieties across three geographic regions.}
	\label{tab:variety_region_data}
	\renewcommand{\arraystretch}{1.4}
	
	\begin{tabular}{|c|c|c|c|}
		\hline
		\multirow{2}{*}{\begin{tabular}{c}
				Variety \\
				(Factor - $A$)
					\end{tabular}}
		& \multicolumn{3}{c|}{Geographic Region (Factor-$B$)} \\
		\cline{2-4}
		& Region-A & Region-B & Region-C \\
		\hline
		
		Variety-W
		& \begin{tabular}{l}
			6.9; 11.8; 6.2;\\
			9.2; 9.2; 6.2
		\end{tabular}
		
		& \begin{tabular}{l}
			8.9; 9.2; 5.2;\\
			7.7; 7.8; 5.7
		\end{tabular}
	
		& \begin{tabular}{l}
			6.8; 5.2; 5.0; \\
			5.2; 5.5; 7.3 
		\end{tabular}  \\
		\hline
		Variety-X
		& \begin{tabular}{l}
			11.0; 7.8; 7.3;\\
			9.1; 7.9; 6.9
		\end{tabular}
		
		& \begin{tabular}{l}
			5.8; 5.1; 5.0;\\
			9.4; 8.3; 5.7
		\end{tabular}
		
		& \begin{tabular}{l}
			7.8; 6.5; 7.0; \\
			9.3; 6.6; 10.8 
		\end{tabular}  \\
		\hline
		Variety-Y
		& \begin{tabular}{l}
			13.1; 12.1; 9.9;\\
			12.4; 11.3; 11.0
		\end{tabular}
		
		& \begin{tabular}{l}
			12.1; 7.1; 13.0;\\
			13.7; 12.9; 7.5
		\end{tabular}
		
		& \begin{tabular}{l}
			8.7; 10.5; 10.0; \\
			8.1; 10.6; 10.5 
		\end{tabular}  \\
		\hline
		Variety-Z
			& \begin{tabular}{l}
			13.4, 14.1, 13.5,\\
			 13.0, 12.3, 13.7
		\end{tabular}
		
		& \begin{tabular}{l}
			9.1, 13.1, 13.2,\\
			8.6, 9.8, 9.9
		\end{tabular}
		
		& \begin{tabular}{l}
			11.8, 13.5, 14.0, \\
			10.8, 12.3, 14.0 
		\end{tabular}  \\
		\hline
	\end{tabular}	
\end{table}	
	
First, let us go ahead and analyse the datasets of Table 1.1 - Table 1.8 using the standard $F-$test under the usual two-way ANOVA assumptions as described in Subsection 1.1. The Table 1.9 summarizes these results which will come handy for comparisons with the results obtained from the proposed new approach based on the gamma model.

The purpose of this work is to show an alternative approach to analyze nonnegative two-factor data based on gamma distributions thereby replacing the homoscedastic normality assumption which brings more robustness to the analysis of two-factor data.

\begin{table}[H]
	\centering
	\caption{Two-way ANOVA results under the usual normal model (significance level $\alpha=0.05$)}
	\renewcommand{\arraystretch}{1.4}
	\begin{tabular}{|l|c|c|c|p{5cm}|}
		\hline
		\multirow{2}{*}{Dataset}
		& \multicolumn{3}{c|}{Source of Variation and $p$-value}
		& \multirow{2}{*}{Conclusion based on ANOVA} \\
		\cline{2-4}
		& $P_{\mathrm{ANOVA}}^{A}$
		& $P_{\mathrm{ANOVA}}^{B}$
		& $P_{\mathrm{ANOVA}}^{AB}$
		& \\ \hline
		
		Example 1.1
		& $<0.001$
		& $<0.001$
		& $0.789$
		&
		\begin{tabular}[c]{@{}l@{}}
			$A$: significant\\
			$B$: significant\\
			$AB$: insignificant
		\end{tabular}
		\\ \hline
		
		Example 1.2
		& $<0.001$
		& $<0.001$
		& $0.023$
		&
		\begin{tabular}[c]{@{}l@{}}
			$A$: significant\\
			$B$: significant\\
			$AB$: significant
		\end{tabular}
		\\ \hline
		
		Example 1.3 	
		& $<0.001$
		& $<0.001$
		& $<0.001$
		&
		\begin{tabular}[c]{@{}l@{}}
			$A$: significant\\
			$B$: significant\\
			$AB$: significant
		\end{tabular}
		\\ \hline
		Example 1.4 	
		& $<0.001$
		& $<0.001$
		& $<0.001$
		&
		\begin{tabular}[c]{@{}l@{}}
			$A$: significant\\
			$B$: significant\\
			$AB$: significant
		\end{tabular}
		\\ \hline
		Example 1.5 	
		& $0.641$
		& $0.565$
		& $0.721$
		&
		\begin{tabular}[c]{@{}l@{}}
			$A$: insignificant\\
			$B$: insignificant\\
			$AB$: insignificant
		\end{tabular}
		\\ \hline
		Example 1.6 	
		& $0.278$
		& $<0.001$
		& $<0.001$
		&
		\begin{tabular}[c]{@{}l@{}}
			$A$: insignificant\\
			$B$: significant\\
			$AB$: significant
		\end{tabular}
		\\ \hline
			Example 1.7 	
		& $0.008$
		& $<0.001$
		& $0.007$
		&
		\begin{tabular}[c]{@{}l@{}}
			$A$: significant\\
			$B$: significant\\
			$AB$: significant
		\end{tabular}
	\\ \hline
		Example 1.8 	
	& $<0.001$
	& $0.006$
	& $0.091$
	&
	\begin{tabular}[c]{@{}l@{}}
		$A$: significant\\
		$B$: significant\\
		$AB$: insignificant
	\end{tabular}
	\\ \hline
	\end{tabular}
	\label{tab:anova_summary}
\end{table}
\subsection{Residual Analysis of the Above Examples under Normality}
Let us now verify the standard assumptions of normality and homoscedascity using the observed residuals for all the eight above datasets (i.e., Example 1.1 - 1.8) as shown in Table~\ref{tab:anova_summary}. To test normality, we employ the two most popular methods, namely - `Shapiro-Wilk test' (SWT) and `Anderson-Darling test' (ADT). On the other hand, for testing homoscedasticity, we use `Barlett's test'  (BT), `Levene's test' (LT) and `Modified Levene's test' (MLT). When normality is guaranteed, then `Barlett's test' is possibly the best method to check homoscedasticity. Otherwise, LT and/or even better the MLT, is considered most rebust. However, it should be kept in mind that these test results are not very reliable due to small sample sizes.

	\begin{table}[htbp]
		\centering
		\caption{Checking model assumptions for usual two-factor ANOVA ($\alpha=0.05$).}
		\label{tab:anova_assumptions}
		\begin{tabular}{|l|r|r|r|r|r|r|}
			\hline
			\multirow{3}{*}{Dataset} & \multicolumn{2}{c|}{Normality} & \multicolumn{3}{c|}{Homoscedasticity} & \multicolumn{1}{c|}{Conclusions} \\ \cline{2-6}
			& \multirow{2}{*}{SWT} & \multirow{2}{*}{ADT} & \multirow{2}{*}{BT} & \multirow{2}{*}{LT} & \multirow{2}{*}{MLT} & \multicolumn{1}{c|}{based on} \\ 
			& & & & & & \multicolumn{1}{c|}{ANOVA} \\ \hline
			
			Example 1.1 & 0.302 & 0.416 & 0.427 & 0.031 & 0.769 & Normality: retained; \\ 
			& & & & & & Homoscedasticity: retained. \\ \hline
			
			Example 1.2 & 0.102 & 0.058 & 0.014 & 0.058 & 0.148 & Normality: Questionable,\\
			& & & & & &  but marginally retained; \\ 
			& & & & & & Homoscedasticity: Questionable,  \\
			& & & & & &  but marginally retained.	\\ \hline
			
			Example 1.3 & 0.898 & 0.634 & 0.872 & 0.568 & 0.931 & Normality: retained; \\ 
			& & & & & & Homoscedasticity: retained. \\ \hline
			
			Example 1.4 & 0.009 & 0.016 & 0.097 & 0.049 & 0.075 & Normality: rejected; \\ 			
			& & & & & & Homoscedasticity: retained. \\ \hline
		
			Example 1.5 & 0.918 & 0.886 & 0.271 & 0.235 & 0.260 & Normality: retained; \\
			& & & & & & Homoscedasticity: retained. \\ \hline
			
			Example 1.6 & 0.699 & 0.948 & 0.657 & 0.548 & 0.758 & Normality: retained; \\ 
			& & & & & & Homoscedasticity: retained. \\ \hline
			Example 1.7 & 0.660 & 0.407 & 0.726 & 0.615 & 0.788 & Normality: retained; \\ 
			& & & & & & Homoscedasticity: retained. \\ \hline
			Example 1.8 & 0.478 & 0.300 & 0.134 & 0.001  & 0.493 & Normality: retained; \\ 
			& & & & & & Homoscedasticity: Questionable,  \\
			& & & & & & but marginally retained.	\\ \hline
		\end{tabular}

	\vspace{0.5 cm}
 	\end{table}
\noindent\textbf{Remark 1.1} Note that all the computed p-values in Tables 1.9 and 1.10 may not be accurate due to very small sample sizes whereas all the usual primary tests are based on exact homoscedastic normality assumptions, and all the diagnostic tests are asymptotic in nature.  
	
\subsection{A New Gamma Model Based Approach}
		
As motivated by the above examples, we propose a break from the regular normal model based two-way ANOVA, and propose an alternative analysis based on the gamma model as follows.

 A nonnegative random variable $X$ is said to have a gamma model with shape parameter $\delta$ and scale parameter $\beta$ (henceforth denoted as $G(\delta,\beta)$ model) provided its pdf is given as
\begin{equation}
f(x\mid\delta,\beta)=\lbrace\Gamma(\delta)\beta^\delta \rbrace^{-1} e^{-x/\beta}x^{\delta-1},
\end{equation}
where $x>0,\delta>0$ and $\beta>0.$ Note that the shape of the distribution is nicely controlled by the shape parameter $\delta$, as $G(\delta,\beta)$ goes from extremely positively skewed (for small $\delta$) to almost symmetric (for large $\delta$), and hence almost normality comes under the versatile gamma model, without the linear additivity assumption of (1.1). (To be precise, the $G(\delta,\beta)$ distribution and the $N(\delta\beta, \delta\beta^2)$ distribution
are very close when $\delta$ is ``large", by the Central Limit Theorem.)

We are going to assume that the observations $X_{ijk}\hspace{0.1cm}(1\le k\le n_{ij})$ are \textit{iid} $G(\delta_{ij},\beta_{ij}),1\le i\le a$ and $1\le j\le b.$ Note that the mean and variance of $X_{ijk}$ under $G(\delta_{ij},\beta_{ij})$ model are
\begin{equation}
E(X_{ijk})=\mu_{ij}\textrm{(say)}=\delta_{ij}\beta_{ij}, \textrm { and}\hspace{0.15cm} V(X_{ijk})=\sigma_{ij}^2\textrm{(say)}=\delta_{ij}\beta_{ij}^2.
\end{equation}

To investigate whether the factor levels, individually or jointly have any significant influence on the means, or whether the joint effect (i.e., the interaction) can be explained in the multiplicative form of the main effects, we propose the following four initial hypothesis testing problems.\\

\textbf{Problem - 1}: Test $H_0^{(1)}:  \mu_{ij} = \mu_{i\cdot}\hspace{0.1cm}\forall \hspace{0.1cm} j$ 
vs. $H_A^{(1)}:\mu_{ij} \neq \mu_{ij'},$\textrm{ for some}\ \ $ j \neq j',1\le j, j'\le b.$

\textbf{Problem - 2}: Test $H_0^{(2)}:\mu_{ij} = \mu_{\cdot j}\hspace{0.1cm}\forall \hspace{0.1cm}i$ vs. $H_A^{(2)}:\mu_{ij} \neq \mu_{i'j},$ \textrm{for some}\ \  $i \neq i', 1\le i, i'\le a.$

\textbf{Problem - 3}: Test $H_0^{(3)}:\mu_{ij} = \mu_{\cdot\cdot}\forall\hspace{0.1cm} (i,j)$ vs. $H_A^{(3)}:\mu_{ij} \neq \mu_{i'j'},$ \textrm{for some} $i \neq i'$\textrm{ and/or } $j\neq j',1\le i, i'\le a, 1\le j, j'\le b$.

\textbf{Problem - 4}: Test $H_0^{(4)}:\mu_{ij}=\mu_{i\cdot}\times\mu_{\cdot j} \hspace{0.1cm}\forall\hspace{0.1cm}(i,j)$  vs. $H_A^{(4)}:\mu_{ij}\neq\mu_{i\cdot}\times\mu_{\cdot j}$ for some combinations of $(i,j)$.

If we fail to reject the null hypothesis:\\
(i) in Problem - 1, then it implies that Factor-$B$ has no influence on the mean response;\\
(ii) in Problem - 2, then it implies that Factor-$A$ has no influence on the mean response;\\ 
(iii) in Problem - 3, then both the factors have no influence on the mean response when they act simultaneously;\\
(iv) in Problem - 4, the joint interaction effect of the two factors is of multiplicative nature. This implies that the change in mean response with respect to the change in one factor level, when the other factor level is held constant, is proportional to the change of mean response of the varying factor. This Problem - 4 null also justifies an approximate linear additive model under the logarithmic transformation.

 Further note that the Problems 1 and 2 are technically identical if one switches the roles of $A$ and $B$. Therefore, we will not show the details of solving Problem - 2, even though the analogous results of the corresponding data analysis will be presented.

The assumption of the equality of scale parameters (i.e., $\beta_{ij}\textrm{'}s$), though heuristic when the factor levels are of similar nature, like the ones in Example 1.3, still needs to be formally tested. Hence we propose an additional hypothesis testing problem as follows:

\textbf{Problem - 5}: Test $H_0^{(5)}:  \beta_{ij} = \beta\hspace{0.1cm}\forall \hspace{0.1cm} (i,j)$ 
vs. $H_A^{(5)}:\beta_{ij}\textrm{'}s$\textrm{ are not identical for all} $(i,j)$.\\

Therefore, given a dataset, we propose the following simple flowchart (see Figure 1.1) for arriving at a logical conclusion.

\begin{figure}[h]
	\centering
	
	\begin{tikzpicture}[
		>=Stealth,
		box/.style={
			draw,
			rectangle,
			minimum height=9mm,
			align=center,
			font=\normalsize
		}
		]
		
		\node[box] (A) at (0,0) {Two-factor dataset under gamma model};
		
		\node[box] (B) at (0,-1.8) {Address Problem - 5\\ (Test $\beta_{ij}=\beta$)};
		
		\node[box] (C) at (-4,-4.5) {Retain $H_0^{(5)}$};
		
		\node[box] (D) at (4,-4.5) {Reject $H_0^{(5)}$};
		
		\node[box,text width=5.8cm] (E) at (-4,-7.3)
		{Address the Problems 1 - 4\\
			under the assumption $\beta_{ij}=\beta$.};
		
		\node[box,text width=5.8cm] (F) at (4,-7.3)
		{Address the Problems 1 - 4\\
			under unequal $\beta_{ij}$'s.};
		
		\draw[-Stealth] (A) -- (B);
		
		\coordinate (J) at (0,-3.0);
		
		\draw[-Stealth] (B) -- (J);
		
		\draw (J) -- (-4,-3.0);
		\draw (J) -- (4,-3.0);
		
		\draw[-Stealth] (-4,-3.0) -- (C);
		\draw[-Stealth] (4,-3.0) -- (D);
		
		\draw[-Stealth] (C) -- (E);
		\draw[-Stealth] (D) -- (F);
		
	\end{tikzpicture}
	
	\caption{A flowchart of analyzing two-way datasets under the gamma model.}
	\label{fig:flowchart_gamma}
	
\end{figure}
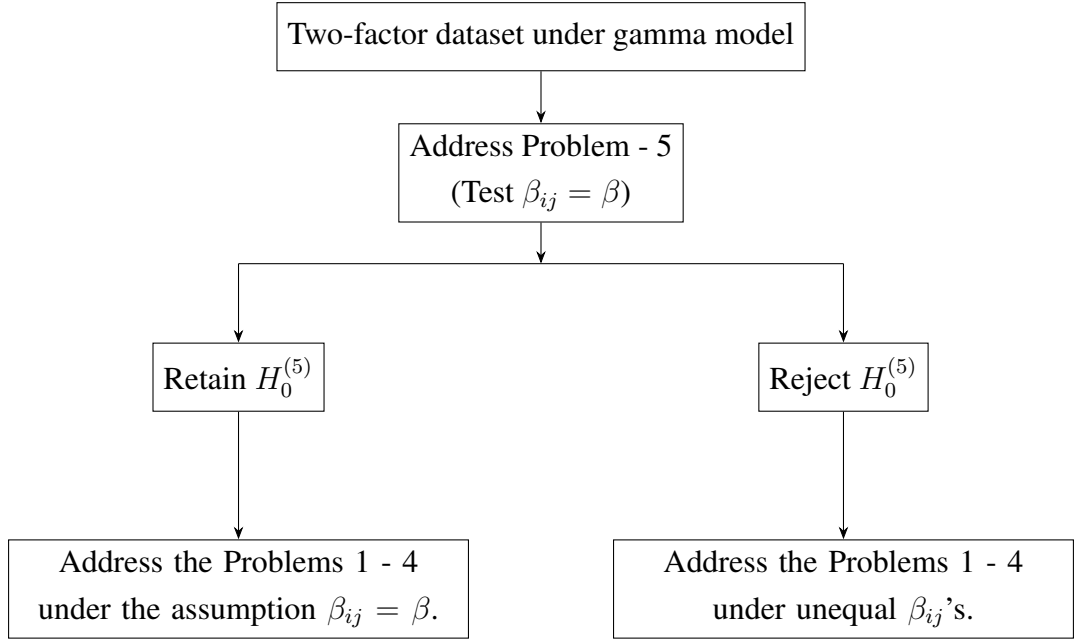

\begin{table}[ht]\caption{Further clarification of the flowchart Figure 1.1 based on the outcome of Problem - 5.}
\fontsize{8pt}{12pt}	
	\begin{tabular}{|l|l||l|l|}\hline
		 \multicolumn{2}{|c||}{Address Problems 1 - 4 when $\beta_{ij}=\beta$ (Case-1)}& \multicolumn{2}{c|}{Address Problems 1 - 4 when $\beta_{ij}\neq\beta$ (Case-2)}\\
	     \hline 
		\multicolumn{2}{|l||}{\textbf{(1)} $H_0^{(11)}: \mu_{ij}=\mu_{i\cdot}$ vs $H_A^{(11)}:\mu_{ij}\neq\mu_{ij'}$ for}
		 &\multicolumn{2}{l|}{\textbf{(1)} $H_0^{(12)}: \mu_{ij}=\mu_{i\cdot}$ vs $H_A^{(12)}:\mu_{ij}\neq\mu_{ij'}$ for}\\
	    \multicolumn{2}{|l||}{some $j\neq j'$}
		&\multicolumn{2}{l|}{some $j\neq j'$}\\
		 Get the global  
		&under $H_0^{(11)}$, get the 
		& Get the global 
		&under $H_0^{(12)},$ get the \\
		 MLEs $\hat\beta,\hat \delta_{ij}$ with   & restricted MLEs & MLEs $\hat \beta_{ij},\hat \delta_{ij} $ with & restricted MLEs\\
		 no restriction & $\hat\delta_{i\cdot}^0,\hat\beta^0$ 
		&  no restriction & $\hat \delta_{ij}^0,\hat \mu_{i\cdot},$
		$\implies \hat \beta_{ij}^0=\hat \mu_{i\cdot}/\hat \delta_{ij}^0$
		\\ \hline
		
		\multicolumn{2}{|l||}{\textbf{(2)} $H_0^{(21)}: \mu_{ij}=\mu_{\cdot j}$ vs $H_A^{(21)}:\mu_{ij}\neq\mu_{ij'}$ for} &\multicolumn{2}{l|}{\textbf{(2)} $H_0^{(22)}: \mu_{ij}=\mu_{\cdot j}$ vs $H_A^{(22)}:\mu_{ij}\neq\mu_{i'j}$ for}\\
		\multicolumn{2}{|l||}{some $i\neq i'$}
		&\multicolumn{2}{l|}{some $i\neq i'$}\\
		
		Get the global 
		 &under $H_0^{(21)}$, get the & Get the global  & under $H_0^{(22)},$ get the \\
		MLEs $\hat\beta,\hat \delta_{ij}$ with & restricted MLEs
		& MLEs $\hat\beta_{ij},\hat \delta_{ij}$ with
		& restricted MLEs\\
		no restriction & $\hat \delta_{\cdot j}^0,\hat\beta^0$ & no restriction & $\hat \delta_{ij}^0,\hat\mu_{\cdot j}$
		$\implies \hat \beta_{ij}^0=\hat \mu_{\cdot j}/\hat \delta_{ij}^0$ \\
		\hline
		 	
		\multicolumn{2}{|l||}{\textbf{(3)} $H_0^{(31)}: \mu_{ij}=\mu_{\cdot\cdot}$ vs $H_A^{(31)}:\mu_{ij}\neq\mu_{i'j'}$ for} &\multicolumn{2}{l|}{\textbf{(3)} $H_0^{(32)}: \mu_{ij}=\mu_{\cdot\cdot}$ vs $H_A^{(32)}:\mu_{ij}\neq\mu_{i'j'}$ for}\\
		\multicolumn{2}{|l||}{some $i\neq i'$ and/or$\hspace{0.1cm}j\neq j'$}
		&\multicolumn{2}{l|}{some $i\neq i'$ and/or$ \hspace{0.1cm}j\neq j'$}\\
		
		Get the global
		  &under $H_0^{(31)}(\delta_{ij}=\delta )$,  & Get the global & under $H_0^{(32)},$ get the \\
	     MLEs $\hat\beta,\hat \delta_{ij}$ with & get the restricted MLEs & MLEs $\hat\beta_{ij},\hat \delta_{ij}$ with & restricted MLEs\\
	     no restriction &  $\hat\beta^0, \hat \mu_{\cdot\cdot},(\hat \delta_{ij}^0=\hat \mu_{\cdot\cdot}/\hat \beta^0)$ 
		 & no restriction
		 & $\hat \delta_{ij}^0,\hat\mu_{\cdot\cdot}$ 
		$\implies\hat \beta_{ij}^0=\hat \mu_{\cdot\cdot}/\hat \delta_{ij}^0$\\
		\hline

		\multicolumn{2}{|l||}{\textbf{(4)} $H_0^{(41)}: \mu_{ij}=\mu_{i\cdot}\times\mu_{\cdot j}$ vs } &\multicolumn{2}{l|}{\textbf{(4)} $H_0^{(42)}: \mu_{ij}=\mu_{i\cdot}\times\mu_{\cdot j}$ vs }\\
		\multicolumn{2}{|l||}{$\hspace{0.7cm} H_A^{(41)}:\mu_{ij}\neq\mu_{i\cdot}\times\mu_{\cdot j}$ for some $(i,j)$}
		&\multicolumn{2}{l|}{ $\hspace{0.7cm} H_A^{(42)}:\mu_{ij}\neq\mu_{i\cdot}\times\mu_{\cdot j}$for some $(i,j)$}\\
		
		Get the global  &under $H_0^{(41)}(\delta_{ij}=\delta_{i\cdot}\times\delta_{\cdot j}), $ 
		& Get the global &under $H_0^{(42)},$ get the \\
		MLEs $\hat\beta,\hat \delta_{ij}$ with
		& get the restricted MLEs &
		MLEs $\hat\beta_{ij},\hat \delta_{ij}$ with & 
		restricted MLEs $\hat \mu_{i\cdot},\hat\mu_{\cdot j},\hat\beta_{ij}^0$  \\
		no restriction & $\delta_{i\cdot},\delta_{\cdot j},\hat\beta^0$
		& no restriction &
		 $\implies\hat \delta_{ij}^0=(\hat \mu_{i\cdot}\times\hat \mu_{\cdot j})/\hat \beta_{ij}^0$\\
		\hline
	
	\end{tabular}
\end{table}
The rest of the paper is organized as follows. In Section 2, we first present the details of addressing Problem - 5 (i.e., testing $H_0^{(5)}$ first). In Section 3, we address the Problem - 1 (i.e., testing $H_0^{(1)}$) under both the cases, when $\beta_{ij}\textrm{'}s$ are all equal (to $\beta$), and when $\beta_{ij}\textrm{'}s$ are not all equal. As stated earlier, we will skip the details of addressing Problem - 2 since technically it is the same as Problem - 1. In Section 4, we address Problem - 3 under both the cases (as in Section 3). Section 5 deals with addressing Problem - 4 in a similar manner. Each section provides extensive simulation results to show the efficacy of our methods. The likelihood ratio test (LRT) statistic has been used to show the performance of the asymptotic LRT (ALRT) test, which is not very accurate in terms of maintaining the nominal level, and a parametric bootstrap (PB) version of the test based on the LRT statistic (henceforth, PBLRT) has been proposed which works remarkably well. Finally, in Section 6 we revisit all the eight examples discussed earlier (in Section 1) under the proposed gamma model and compare the outcomes with those under the usual normality assumption. 

\noindent\textbf{Remark 1.2} The whole idea behind this work is to present a different perspective when the nonnegative observations are 
generated under two potential factors, using a more general gamma model, thereby providing a more robust analysis of the data. We sincerely believe that the proposed new approach would be extremely beneficial to the applied researchers dealing with small samples of nonnegative observations. Furthermore, the R package code for our simulation studies, as well as the implementations of the examples discussed herein, have been made available in the Appendix.

\section{\fontsize{15.8pt}{16pt}\selectfont Testing the equality of scale parameters (Problem - 5)}
Our goal in this section is to consider the Problem - 5 first with the hypothesis testing of common scale, i.e., $H_0^{(5)}:\beta_{ij}=\beta\hspace{0.2cm}\forall\hspace{0.1cm}(i,j),$ vs. $H_A^{(5)}: \beta_{ij}$ are not equal for some $(i,j)$.

Given the independent observations $X_{ijk}\sim$ $G(\delta_{ij},\beta_{ij})$, the likelihood function L is given as
\begin{align}
L = L(\delta_{ij},\beta_{ij},1\le i\le a,1\le j\le b|X_{ijk},\forall\hspace{0,1cm}(i,j,k)) \nonumber \\
 =\prod_{i=1}^a\prod_{j=1}^b\prod_{k=1}^{n_{ij}}\left[\{\Gamma(\delta_{ij})\beta_{ij}^{\delta_{ij}}\}^{-1}e^{-X_{ijk}/\beta_{ij}}(X_{ijk})^{\delta_{ij}-1}\right].
\end{align}
Thus, the log-likelihood function $L_* = ln L$ can be written as
\begin{align}
L_* = \sum_{i=1}^a\sum_{j=1}^b \bigg\{- n_{ij}ln\Gamma(\delta_{ij})- n_{ij}\delta_{ij}ln\beta_{ij} -  
(1/\beta_{ij})\sum_{k=1}^{n_{ij}}X_{ijk}+(\delta_{ij} - 1)\sum_{k=1}^{n_{ij}}ln{X_{ijk}} \bigg\}.
\end{align}
We use the notation $\bar X_{ij\cdot}$ and $\widetilde X_{ij\cdot}$ to denote the arithmetic mean (AM) and geometric mean (GM) of the observations in the $(i,j)^{th}$ cell, i.e.,
\begin{equation}
\bar X_{ij\cdot} = \sum_{k=1}^{n_{ij}}X_{ijk}/n_{ij};\hspace{0.1cm} \widetilde X_{ij\cdot}=(\prod_{k=1}^{n_{ij}}X_{ijk})^{1/n_{ij}}.
\end{equation}
Then $L_*$ can be simplified as
\begin{align}
&L_* = \sum_{i=1}^a\sum_{j=1}^b n_{ij}\lbrace - ln\Gamma(\delta_{ij})-\delta_{ij}ln\beta_{ij} - (1/\beta_{ij})\bar X_{ij\cdot} 
+ (\delta_{ij}-1)ln \widetilde X_{ij\cdot}\rbrace.
\end{align}
By differentiating $L_*$ in (2.4) w.r.t $\delta_{ij}$ and $\beta_{ij},$ and then setting them equal to zero yields the following system of equations
\begin{equation}
\psi(\hat\delta_{ij})+ln(\hat\beta_{ij})=ln(\tilde X_{ij\cdot});\text{ and } \hat\beta_{ij}=\bar X_{ij\cdot}/\hat\delta_{ij}, \hspace{0.1cm}\forall\hspace{0.1cm}(i,j),
\end{equation}
where $\psi(c)=\lbrace\partial ln\Gamma(c)/\partial c\rbrace$ is the di-gamma function defined at $c>0.$
Futher simplication yields
\begin{equation}
\psi(\hat\delta_{ij})-ln(\hat\delta_{ij})=ln(\tilde X_{ij\cdot}/\bar X_{ij\cdot}).
\end{equation}
 We can find the values of $\hat\delta_{ij},\hat\beta_{ij}$, the MLEs of $\delta_{ij}$ and $\beta_{ij}$ respectively, by solving (2.5) and (2.6) with the total number of equations being $(2ab)$.
 
 On the other hand, under $H_0^{(5)}:\beta_{ij}=\beta,\hspace{0.1cm}\forall\hspace{0.1cm} (i,j)$, the restricted MLEs of $\beta$ and $\delta_{ij}$ (denoted by $\hat\beta^{0}$ and $\hat\delta_{ij}^0$, respectively) are found as the following: First get $\hat \delta_{ij}^0$ by solving
\begin{equation}
\psi(\hat\delta_{ij}^0)-ln(\sum_{i_0=1}^a\sum_{j_0=1}^b n_{i_0j_0}\hat\delta_{i_0j_0}^0)=ln(\widetilde X_{ij\cdot}\Big/\sum_{i_0=1}^a\sum_{j_0=1}^b n_{i_0j_0}\bar X_{i_0j_0\cdot}).
\end{equation}
Then get $\hat\beta^{0}$ by
\begin{equation}
\hat\beta^0=(\sum_{i=1}^a\sum_{j=1}^b n_{ij}\bar X_{ij\cdot}) \Big/ (\sum_{i=1}^a\sum_{j=1}^b n_{ij}\hat\delta_{ij}^0).
\end{equation}
So, our likelihood ratio test (LRT) statistic is $\Lambda_*^{(5)}=-2\ln\Lambda^{(5)}$, where
\begin{equation}
\Lambda^{(5)}=\frac{L(\hat\delta_{ij}^0,\hat\beta^0\mid \textrm { data})}{L(\hat\delta_{ij},\hat\beta_{ij}\mid \textrm { data})}=\frac{L(\hat\delta_{ij}^0,\hat\beta^0\mid X_{ijk} \hspace{0.1cm}\forall \hspace{0.1cm} i,j,k)}{L(\hat\delta_{ij},\hat\beta_{ij}\mid X_{ijk}\hspace{0.1cm}\forall\hspace{0.1cm} i,j,k)}.
\end{equation}
ALRT says that under $H_0^{(5)},\Lambda_*^{(5)}\sim\chi_\nu^2 $  approximately, where $\nu=2ab-(ab+1)=ab-1.$ So, reject $H_0^{(5)}$ if $\Lambda_*^{(5)}>\chi_{\nu,\alpha}^2.$

As we stated earlier and shown in Table 2.1, for small $n_{ij}$ values, the size of the ALRT is much higher than $\alpha$, whereas the proposed PBLRT described below keeps it very close to $\alpha$. Before seeing these results, now we describe the steps of the proposed PBLRT, details of which can be found in Chang et al. (2008, 2010). The one-factor analysis of data under the gamma model (i.e., "ANOGaM-1") was considered by Chang et al. (2011), but the current analysis gamma model under two factors (i.e., ANOGaM-2) is far more 
complex and computationally challenging than the one-factor version.
\\

\noindent{\bf{Steps of the Test Based on Parametric Bootstrap (PB) Using the LRT Statistic (or PBLRT):}}

\textbf{Step 1}. Calculate $\Lambda_*$ from the original data $\{X_{ijk}, \forall (i,j,k)\}$.

\textbf{Step 2}. Assume that $H_0^{(5)}$ is true, i.e., $\beta_{ij}=\beta.$ Then generate a bootstrap sample $X_{ijk}^{(m)}$ in the $m^{th}$ replication, where $X_{ijk}^{(m)}\sim G(\hat\delta_{ij}^0,\hat\beta^0)$ with the ranges of $i,j$ and $k$ remain as before. With these replicated observations, get the unrestricted as well as restricted MLEs as $\hat \delta_{ij}^{(m)},\hat \beta_{ij}^{(m)},\hat \delta_{ij}^{0(m)},\hat \beta^{0(m)},$ and recompute $\Lambda_*$ as $\Lambda_*^{(m)}.$

\textbf{Step 3}. Repeat the above Step 2 a large number of times $m=1,2,3,...,M,$ and get $\Lambda_*^{(1)},\Lambda_*^{(2)},...,\Lambda_*^{(M)}.$

\textbf{Step 4}. Order $\Lambda_*^{(1)},\Lambda_*^{(2)},...,\Lambda_*^{(M)}$ as $\Lambda_{*(1)}\leq\Lambda_{*(2)}\leq...\leq\Lambda_{*(M)}.$

\textbf{Step 5}. With the cut-off point of PBLRT as $\Lambda_{*u}=\Lambda_{*((1-\alpha)M)},$ reject $H_0^{(5)}$ if $\Lambda_*>\Lambda_{*u};$ and retain $H_0^{(5)}$ otherwise. Alternatively, one can compute the $p$-value of the proposed $PBLRT$ as $\alpha_{PBLRT}\approx\sum_{m=1}^M I(\Lambda_{*(m)}>\Lambda_*)/M.$ 

To see how the ALRT and PBLRT perform in term of P(Type-I error), we are going to run simulations to get approximate sizes. Replicate each test $Q$ times and observe the proportion of times (out of $Q$) a test rejects $H_0^{(5)}$.

 For all $q=1,2,...,Q,$ we define  $I_{ALRT}^{(q)}=1$ if ALRT rejects $H_0^{(5)}, I_{ALRT}^{(q)}=0,$ otherwise; and $I_{PBLRT}^{(q)}=1$ if PBLRT rejects $H_0^{(5)}, I_{PBLRT}^{(q)}=0,$ otherwise. Depending on the input parameters, we define the size of ALRT and PBLRT, denoted by $\alpha_{ALRT}$ and $\alpha_{PBLRT}$, respectively, as $\alpha_{ALRT}\approx\sum_{q=1}^Q I_{ALRT}^{(q)}/Q$ and  $\alpha_{PBLRT}\approx \sum_{q=1}^Q I_{PBLRT}^{(q)}/Q.$ 
 
Without loss of generality, take $\beta_{ij}=\beta = 1, a = b = 2,$ the simulated results of the Problem - 5 as shown in the following Table 2.1. The patterns remains the same for other combinations of $(a,b)$ as well.

\begin{table}[H]
	\caption{Simulated size of the two tests for Problem - 5, $a=b=2,\beta=1.$}
	\centering	
		\small
			\begin{tabular}{|*{8}{c|}}\hline
				\multirow{2}{*}{$n_{ij}$}&\multirow{2}{*}{$\delta_{ij}$}&\multicolumn{2}{c|}{$\alpha=0.01$}&\multicolumn{2}{c|}{$\alpha=0.05$}&\multicolumn{2}{c|}{$\alpha=0.10$}\\
				\cline{3-8}
			&&\multicolumn{1}{c|}{$\alpha_{ALRT}^{(5)}$}&{$\alpha_{PBLRT}^{(5)}$}&{$\alpha_{ALRT}^{(5)}$}&{$\alpha_{PBLRT}^{(5)}$}&{$\alpha_{ALRT}^{(5)}$}&{$\alpha_{PBLRT}^{(5)}$}\\
			\hline
			\multirow{5}{*}{$5$}&$(2,2,2,2)$&0.046&0.014&0.131&0.051&0.212&0.104\\
			\cline{2-8}
			&$(5,5,5,5)$&0.044&0.012&0.128&0.051&0.211&0.100\\
			\cline{2-8}
			&$(5,5,10,10)$&0.047&0.012&0.127&0.057&0.204&0.106\\\cline{2-8}
			&$(5,10,10,10)$&0.040&0.010&0.130&0.050&0.215&0.103\\
			\cline{2-8}
			&$(10,10,10,10)$&0.034&0.008&0.129&0.045&0.209&0.098\\\hline
			\multirow{5}{*}{$10$}&$(2,2,2,2)$&0.024&0.009&0.079&0.052&0.151&0.103\\
			\cline{2-8}	
			&$(5,5,5,5)$&0.025&0.011&0.089&0.059&0.153&0.108\\
			\cline{2-8}
			&$(5,5,10,10)$&0.022&0.011&0.085&0.053&0.154&0.103\\\cline{2-8}
			&$(5,10,10,10)$&0.019&0.011&0.084&0.055&0.149&0.102\\\cline{2-8}
			&$(10,10,10,10)$&0.019&0.011&0.080&0.045&0.144&0.100\\\hline
			
			\multirow{5}{*}{$25$}&$(2,2,2,2)$&0.013&0.010&0.059&0.046&0.111&0.097\\
			\cline{2-8}	
			&$(5,5,5,5)$&0.012&0.009&0.055&0.047&0.114&0.096\\
			\cline{2-8}
			&$(5,5,10,10)$&0.011&0.008&0.062&0.052&0.121&0.105\\\cline{2-8}
			&$(5,10,10,10)$&0.012&0.010&0.057&0.046&0.113&0.098\\\cline{2-8}
			&$(10,10,10,10)$&0.014&0.011&0.064&0.053&0.119&0.101\\\hline
				
			\multirow{5}{*}{$50$}&$(2,2,2,2)$&0.012&0.011&0.059&0.054&0.113&0.102\\
			\cline{2-8}	
			&$(5,5,5,5)$&0.008&0.007&0.048&0.043&0.101&0.092\\
			\cline{2-8}
			&$(5,5,10,10)$&0.012&0.009&0.057&0.052&0.111&0.102\\\cline{2-8}
			&$(5,10,10,10)$&0.011&0.010&0.053&0.047&0.105&0.098\\\cline{2-8}
			&$(10,10,10,10)$&0.009&0.009&0.047&0.042&0.098&0.089\\\hline
			\end{tabular}

\end{table}
\noindent\textbf{Remark 2.1} Table 2.1 shows that our proposed PBLRT is much superior to ALRT in terms of maintaining the nominal level. For small sample sizes, especially when $n_{ij}$ are less than 25, the size of ALRT is higher than the nominal level by a magnitude ranging from 10\% to 360\%. On the other hand, the size of PBLRT follows the level very closely even for extremely small sample sizes. When the sample sizes get larger, especially near 50, then ALRT is as good as PBLRT in maintaining the level closely. 

Now we apply the above theory of Problem - 5, under the gamma model, for Examples 1.1 - 1.6. For each example, we assume that the data follows suitable gamma models, and then test the equality of scale parameters. The results are summarized in Table 2.2. 

\begin{table}[H]	\caption{Problem - 5 $p$-values for the eight examples and the decisions (using $\alpha=0.05$).}
	\centering
	\begin{tabular}{|c|c||c|c||c|c||}
		\hline
		Dataset & 
		\makecell{$\Lambda_{*}^{(5)}$ } & 
		$P_{ALRT}^{(5)}$  & 
		\makecell{Decision on $H_0^{(5)}$ \\based on ALRT} & 
		$P_{PBLRT}^{(5)}$  & 
		\makecell{Decision on $H_0^{(5)}$\\ based on PBLRT} \\
		\hline
		Example 1.1 & 3.665 & 0.979 & Retain & 0.994 & Retain \\
		\hline
		Example 1.2 & 13.212 & 0.021 & \bf{Reject} & 0.112 & \bf{Retain} \\
		\hline
		Example 1.3 & 2.357 & 0.968 & Retain & 0.998 & Retain \\
		\hline
	    Example	1.4 &13.118   &0.022  & Reject &0.042 & Reject \\
		\hline
		Example 1.5 &9.326   & 0.097  & Retain &0.280 & Retain \\
		\hline
		Example 1.6 &7.529   &0.376  &Retain  &0.485 &Retain  \\
		\hline
		Example 1.7 &7.154   &0.520  &Retain  &0.858 &Retain  \\
		\hline
		Example 1.8 &22.886   &0.018  &\bf{Reject}  &0.081 &\bf{Retain}  \\
		\hline
	\end{tabular}
	\label{tab:ALRT_PBLRT}
\end{table}
\noindent\textbf{Remark 2.2} Interestingly, out of the eight examples, in two cases (Examples 1.2 and 1.8), the two tests give differing inferences where the more reliable PBLRT is indicating the equality of scales. Also, for Example 1.4, both the tests are indicating the inequality of scales, whereas for the remaining five examples they both point toward equality of scales under the gamma model.

\section{Testing the significance of Factor - B (Problem - 1)}
In this section, we consider the Problem - 1: $H_0^{(1)}:\mu_{ij}=\mu_{i\cdot},\forall\hspace{0.1cm}(i,j)$ against the alternative which negates it, under both the cases, when $\beta_{ij}\textrm{'}s$ are all equal and when $\beta_{ij}\textrm{'}s$ are not equal.

\underline{\bfseries{Case-1}}: $\beta_{ij}=\beta$ for all $(i,j).$

Now, we consider the first case, i.e. the scale parameters $\beta_{ij}\textrm{'}s$ are equal (to $\beta$). Then the null hypothesis is stating that the Factor-B has no effect on the mean response, so the null can be written as (since $\mu_{ij}=\beta\delta_{ij}$ under case-1, from (1.4))
\begin{equation}
H_0^{(11)}:\delta_{ij}=\delta_{i\cdot},\forall j, \text{ for some suitable } \delta_{i\cdot;}
\end{equation}
where  $\delta_{i\cdot}$ can be thought as $(\mu_{i\cdot}/\beta).$
The way to test (3.1) is somewhat similar to the one in Section 2. For the given independent observation $X_{ijk}\sim G(\delta_{ij},\beta)$ the likelihood function $L$ is given as
\begin{align}
L=L(\delta_{ij},\beta,1\leq i\leq a,1\leq j \leq b|X_{ijk},\forall\hspace{0.1cm}(i,j,k)) \nonumber \\
=\prod_{i=1}^a\prod_{j=1}^b\prod_{k=1}^{n_{ij}}\left[\{\Gamma(\delta_{ij})\beta^{\delta_{ij}}\}^{-1}e^{-X_{ijk}/\beta}(X_{ijk})^{\delta_{ij}-1}\right]. 
\end{align}
Thus, the log-likelihood function $L_*=\ln L$ can be written as 
\begin{align}
L_* = \sum_{i=1}^a\sum_{j=1}^b \{- n_{ij}ln\Gamma(\delta_{ij})- n_{ij}\delta_{ij}ln\beta -  
(1/\beta)\sum_{k=1}^{n_{ij}}X_{ijk}+(\delta_{ij} - 1)\sum_{k=1}^{n_{ij}}ln{X_{ijk}} \}.
\end{align}
We insert $\bar{X}_{ij\cdot}$ (the arithmetic mean - AM) and $\widetilde{X}_{ij\cdot}$ (the geometric mean - GM) into (3.3), to get
\begin{align}
L_* = \sum_{i=1}^a\sum_{j=1}^b n_{ij}\lbrace - ln\Gamma(\delta_{ij})-\delta_{ij}ln\beta - (1/\beta)\bar X_{ij\cdot} 
+ (\delta_{ij}-1)ln \widetilde X_{ij\cdot}\rbrace.
\end{align}
By differentiating $L_*$ in (3.4) w.r.t $\delta_{ij}$ and $\beta$, and then setting them equal to zero yields the following system of equations
\begin{align}
\psi(\delta_{ij})+ln\beta = ln\widetilde X_{ij\cdot}, \forall \hspace{0.1cm}(i,j); 
\textrm{ and } \quad  \beta(\sum_{i=1}^a\sum_{j=1}^b n_{ij}\delta_{ij}) = \sum_{i=1}^a\sum_{j=1}^b n_{ij}\bar X_{ij\cdot};
\end{align}
Then, solving the system of equations in (3.5), yields the MLEs of $\delta_{ij}$ and $\beta$, say $\hat\delta_{ij}$ and $\hat\beta$, as follows. Essentially, the unrestricted MLEs of the parameters $(\beta,\delta_{ij})$ are those in Problem-5 under the equal scale null restriction.

First obtain $\hat\delta_{ij}$ by solving the following system of $a\times b$ equations
\begin{align}
\psi(\hat\delta_{ij})-ln(\sum_{i_0=1}^a\sum_{j_0 = 1}^bn_{i_0 j_0}\hat\delta_{i_0 j_0}) 
= ln(\widetilde X_{ij\cdot}\big/\sum_{i_0=1}^a\sum_{j_0=1}^b n_{i_0j_0}\bar X_{i_0j_0\cdot}),\forall\hspace{0,1cm} (i,j);
\end{align}
which the same as (2.8) and then obtain  $\hat\beta$ as
\begin{align}
&\hat\beta =(\sum_{i=1}^a\sum_{j=1}^b n_{ij}\bar X_{ij\cdot})\big/(\sum_{i=1}^a\sum_{j=1}^b n_{ij}\hat\delta_{ij}) , 
&
\end{align}
which is again same as (2.9). Thus,
\begin{equation}
\sup L = L(\hat\delta_{ij},\hat\beta,1\le i\le a, 1\le j\le b|X_{ijk},\forall\hspace{0,1cm}(i,j,k)). 
\end{equation}

The log-likelihood function under $H_0^{(11)}$, henceforth denoted by $L_*^{0(11)}$, is
\begin{align}
L_*^{0(11)} = \sum_{i=1}^a\sum_{j=1}^b n_{ij}\lbrace -ln\Gamma(\delta_{i\cdot})-\delta_{i\cdot}ln\beta - (1/\beta)\bar X_{ij\cdot} + (\delta_{i\cdot}-1)ln\widetilde X_{ij.}\rbrace.
\end{align}
Differentiating $L_*^{0(11)}$ w.r.t. $\delta_{i\cdot}$ and $\beta$, and then setting them equal to zero yields
\begin{align}
\sum_{j=1}^b n_{ij}\psi(\delta_{i\cdot})= \sum_{j=1}^b n_{ij}\lbrace ln\widetilde X_{ij\cdot}-\ln\beta\rbrace,\forall\hspace{0.1cm} i; 
\textrm{ and } \quad  \beta\sum_{i=1}^a\sum_{j=1}^b n_{ij}\delta_{i\cdot} = \sum_{i=1}^a\sum_{j=1}^b n_{ij}\bar X_{ij\cdot}
\end{align}
Define the total sample size subject to $i^{th}$ level of Factor-A, and the corresponding sampling proportion as
\begin{equation}
n_{i\cdot}=\sum_{j=1}^b n_{ij}, \textrm { and }\hspace{0.1cm} v_{ij}=n_{ij}/n_{i\cdot}
\end{equation}
The MLEs of $\delta_{i\cdot}$ and $\beta$ under $H_0^{(11)}$, denoted by $\hat\delta_{i\cdot}^0$ \textrm{ and } $\hat\beta^0$, are obtained as follows. First obtain $\hat\delta_{i\cdot}^0$ by solving the following system of $a$ equations
\begin{align}
ln(\sum_{q=1}^a n_{q\cdot}\hat\delta_{q\cdot}^0) -\sum_{j=1}^b v_{ij}\psi(\hat\delta_{i\cdot}^0) 
= ln((\sum_{q=1}^a\sum_{l=1}^b n_{ql}\bar X_{ql\cdot})\big/\prod_{j=1}^b\widetilde X_{ij\cdot}^{v_{ij}}).
\end{align}
and then obtain $\hat\beta^0$ as
\begin{equation}
\hat\beta^0 = (\sum_{i=1}^a\sum_{j=1}^b n_{ij}\bar X_{ij\cdot})\big/(\sum_{i=1}^a\sum_{j=1}^b n_{ij}\hat\delta_{i\cdot}^0).
\end{equation}
Thus, 
\begin{equation}
\sup_{H_0^{(11)}} L = L(\hat\delta_{i\cdot}^0,\hat\beta^0,1\le i\le a|X_{ijk}\hspace{0,2cm} \forall\hspace{0,1cm} (i,j,k)).
\end{equation}
Define the LRT statistic $\Lambda_*^{(11)}=-2\ln\Lambda^{(11)}$, where
\begin{equation}
\Lambda^{(11)}=\frac{L(\hat\delta_{i\cdot}^0,\hat\beta^0\mid \textrm { data})}{L(\hat\delta_{ij},\hat\beta\mid \textrm { data})}=\frac{L(\hat\delta_{i\cdot}^0,\hat\beta^0\mid X_{ijk} \hspace{0.1cm}\forall \hspace{0.1cm} i,j,k)}{L(\hat\delta_{ij},\hat\beta\mid X_{ijk}\hspace{0.1cm}\forall\hspace{0.1cm} i,j,k)}.
\end{equation}

As stated earlier, for `moderately large' $n_{ij}$ values, according to ALRT $\Lambda_*^{(11)}$ follows $\chi_{\nu_{11}}^2$ under $H_0^{(11)},$ with $\nu_{11} = a(b-1).$ We will see later that for small $n_{ij}\textrm{'}s$, the size of the ALRT is higher than $\alpha,$ whereas the proposed PBLRT keeps it almost equal to $\alpha.$ The beauty of the PBLRT is that it is a purely computational technique where one does not need to know the sampling distribution of the test statistic (which is the LRT statistic in this case), and the critical value is derived automatically through simulation. Before discussing further about the pros and cons of the PBLRT, we first describe how it is implemented through a series of steps as given below.\\

\noindent\textbf{Steps of the proposed PBLRT:}

\textbf{Step - 1}: Given the original data $\lbrace X_{ijk},\hspace{0,2cm}\forall\hspace{0,1cm} (i,j,k)\rbrace,$ obtain the unrestricted MLEs $(\hat\delta_{ij},\hat\beta)$ as well as restricted MLEs $(\hat\delta_{i\cdot}^0,\hat{\beta^0})$ (under $H_0^{(11)}).$ Compute $\Lambda_*$ using (3.8) and (3.14).

\textbf{Step - 2}:
 Assuming that $H_0^{(11)}$ is true, generate artificial (bootstrap) observations in an internal loop of $M$ replications. In the $m^{th}$ replications we generate $X_{ijk}^{(m)}$ from Gamma$(\hat\delta_{i\cdot}^0,\hat\beta^0),\hspace{0.1cm} 1\le k\le n_{ij},\hspace{0.1cm} 1\le j\le b,\hspace{0.1cm} 1\le i\le a.$

\textbf{Step - 3}: With the artificial observations $\lbrace X_{ijk}^{(m)},\hspace{0,1cm}\forall\hspace{0.1cm} (i,j,k)\rbrace$ recompute $(\hat\delta_{ij},\hat\beta)$ and $(\hat\delta_{i\cdot}^0,\hat\beta^0)$ as done in Step - 1, and call them $(\hat\delta_{ij}^{(m)},\hat\beta^{(m)})$ and $(\hat\delta_{i\cdot}^{0(m)},\hat{\beta}^{0(m)})$, respectively. Then obtain $\Lambda_*$ value as done in Step - 1, and call it $\Lambda_*^{(m)}$.

\textbf{Step - 4}: By repeating above (i) - (ii) for $m = 1, 2, \cdots, M,$ we have $\Lambda_*^{(1)},\Lambda_*^{(2)},\cdots,\Lambda_*^{(M}).$ Order these $\Lambda_*^{(m)}$ values as $\Lambda_{*(1)}\le \Lambda_{*(2)}\le\cdots\le \Lambda_{*(M)}.$

\textbf{Step - 5}: The critical value for the statistic $\Lambda_*$ (in Step - 1) is obtained as $\Lambda_{*((1-\alpha)M)},$ where $\alpha$ is the level of the test. If $\Lambda_* > \Lambda_{*((1-\alpha)M)},$ then reject $H_0^{(11)}$; retain $H_0^{(11)}$ if otherwise. Alternatively, the p-value of PBLRT is approximated by $\sum_{m=1}^M I(\Lambda_*^{(m)}>\Lambda_*)/M.$

In the following Table 3.1, we show the simulated results of the Problem-1  for case $n_{ij}=n,\forall\hspace{0.1cm}(i,j),1\leq i\leq a,1\leq j\leq b$ and $n$ is taken as $5,10,25,50;~ M=Q=5000;~ a=b=2; ~\alpha =0.01, 0.05 \textrm{ and } 0.10.$
\begin{table}[t] \caption{Simulated size of two tests with $a=b=2,\delta_{1\cdot}= \beta = 1.0 $  (Problem - 1, Case-1).} 	
\fontsize{9pt}{12pt} 
		\begin{tabular}{|*{12}{c|}} \hline 
				\multirow{2}{*}{$\alpha$}&\multirow{2}{*}{$n_{ij}$}&\multicolumn{2}{c|}{$\delta_{2\cdot}=0.5$}&\multicolumn{2}{c|}{$\delta_{2\cdot}=1.0$}&\multicolumn{2}{c|}{$\delta_{2\cdot}=2.0$}&\multicolumn{2}{c|}{$\delta_{2\cdot}=5.0$}&\multicolumn{2}{c|}{$\delta_{2\cdot}=10.0$} \\ \cline{3-12}
				
				&&\multicolumn{1}{c|}{$\alpha_{ALRT}^{(11)}$}&$\alpha_{PBLRT}^{(11)}$&$\alpha_{ALRT}^{(11)}$&$\alpha_{PBLRT}^{(11)}$&$\alpha_{ALRT}^{(11)}$&$\alpha_{PBLRT}^{(11)}$&$\alpha_{ALRT}^{(11)}$&$\alpha_{PBLRT}^{(11)}$&$\alpha_{ALRT}^{(11)}$&$\alpha_{PBLRT}^{(11)}$ \\ \hline
				
				\multirow{4}{*}{$0.01$}
				
&5  &0.021 &0.010 &0.023 &0.010 &0.026 &0.009 &0.028 &0.011 &0.028 &0.011 \\ \cline{2-12}
&10 &0.014 &0.009 &0.016 &0.012 &0.015 &0.011 &0.016 & 0.011 & 0.015 &0.010 \\ \cline{2-12}
&25 &0.011 &0.009 &0.009 &0.008 &0.015 &0.012 &0.013 &0.012 &0.013 &0.010 \\ \cline{2-12}
&50 &0.012 &0.011 &0.009 &0.009 &0.011 &0.011 &0.009 &0.008 &0.011 &0.010 \\ \hline		
				\multirow{4}{*}{$0.05$}
& 5 &0.090 &0.050 & 0.086 &0.047 & 0.090 &0.054 &0.099 &0.057 &0.094 &0.053 \\ \cline{2-12}
&10 &0.065 &0.047 & 0.070 &0.052 &0.064 &0.049 &0.069 &0.054 &0.070 &0.053 \\ \cline{2-12}
&25 &0.053 &0.047 &0.055 &0.051 &0.061 &0.055 &0.059 &0.052 &0.054 &0.048 \\ \cline{2-12}
&50 &0.057 &0.055 &0.051 &0.050 &0.053 &0.051 &0.052 &0.050 &0.052 &0.050 \\ \hline
				
				\multirow{4}{*}{$0.10$}
& 5  &0.153 &0.101 &0.156 &0.097 &0.156 &0.102 &0.167 &0.111 &0.166 &0.104 \\ \cline{2-12}
&10 &0.124 &0.103 &0.125 &0.100 &0.120 &0.097 &0.128 &0.102 &0.129 &0.105 \\ \cline{2-12}
&25 & 0.106 &0.098 &0.110 &0.100 &0.113 &0.105 &0.115 &0.105 &0.111 &0.102 \\ \cline{2-12}
&50 &0.118 & 0.114 &0.102 &0.099 &0.104 &0.099 &0.101 &0.097 &0.105 &0.101 \\ \hline		
\end{tabular}
\end{table}
\\		
\underline{\bfseries{Case-2}}:  $\beta_{ij}\textrm {'}s$ are not equal for some $(i,j)$. 

The log-likelihood function is given as follows,
\begin{align}
L_* = \sum_{i=1}^a\sum_{j=1}^b n_{ij}\{- ln\Gamma(\delta_{ij})-\delta_{ij}ln(\mu_{ij}/\delta_{ij}) -  
(\delta_{ij}/\mu_{ij})\bar X_{ij\cdot}+(\delta_{ij} - 1)ln{\widetilde X_{ij\cdot}} \}.
\end{align}
By differentiating $L_*$ in (3.15) w.r.t $\delta_{ij},$ and $\mu_{ij}$ and then setting them equal to zero yields the following system of equations
\begin{equation}
\psi(\delta_{ij})+ln(\mu_{ij}/\delta_{ij})+\bar X_{ij\cdot}/\mu_{ij}=1+ln\widetilde X_{ij\cdot}
\textrm { and }\beta_{ij}=\bar X_{ij\cdot}/\delta_{ij},\forall\hspace{0.1cm}(i,j).
\end{equation}
Substitute $\beta_{ij}$ in the first equation of (3.17), we get
\begin{equation}
\psi(\delta_{ij})-ln(\delta_{ij})=ln(\widetilde X_{ij\cdot}/\bar X_{ij\cdot}),\forall\hspace{0.1cm}(i,j),
\end{equation}
which is same as (2.7). The log-likelihood function under $H_0^{(12)}:\beta_{ij}=\mu_{i\cdot}/\delta_{ij},1\leq i \leq a,1\leq j \leq b$, henceforth denoted by $L_*^{0(12)}$ is
\begin{align}
L_*^{0(12)} = \sum_{i=1}^a\sum_{j=1}^b n_{ij}\{- ln\Gamma(\delta_{ij})-\delta_{ij}ln(\mu_{i\cdot}/\delta_{ij}) -  
(\delta_{ij}/\mu_{i\cdot})\bar X_{ij\cdot}+(\delta_{ij} - 1)ln{\widetilde X_{ij\cdot}} \}.
\end{align}
Differentiating $L_*^{0(12)}$ w.r.t $\delta_{ij}$ and $\mu_{i\cdot}$, and then setting them equal to zero yields the estimates of $\delta_{ij}$ and $\mu_{i\cdot}$ under $H_0^{(12)},$ denoted by $\hat\delta_{ij}^0$ and $\hat\mu_{i\cdot}$
\begin{equation}
\psi(\hat\delta_{ij}^0)+ln(\hat\mu_{i\cdot}/\hat\delta_{ij}^0)+\bar X_{ij\cdot}/\hat\mu_{i\cdot}=1+ln\widetilde X_{ij\cdot} \textrm{ and } \hat\mu_{i\cdot}=(\sum_{j=1}^b n_{ij} \hat\delta_{ij}^0\bar X_{ij\cdot})\big/(\sum_{j=1}^bn_{ij}\hat\delta_{ij}^0).
\end{equation}
By inserting $\hat\mu_{i\cdot}$ into the first equation of (3.20) we get
\begin{equation}
\psi(\hat\delta_{ij}^0)+ln\{(\sum_{j=1}^bn_{ij}\hat\delta_{ij}^0\bar X_{ij\cdot})\big/(\sum_{j=1}^bn_{ij}\hat\delta_{ij}^0)\}+\bar X_{ij\cdot}\{(\sum_{j=1}^bn_{ij}\hat\delta_{ij}^0)\big/(\sum_{j=1}^b n_{ij}\hat\delta_{ij}^0\bar X_{ij\cdot})\}=1+ln(\hat\delta_{ij}^0\widetilde X_{ij\cdot}).
\end{equation}
Therefore, first we get $\hat\delta_{ij}^0$ by solving (3.21), and then get $\hat\mu_{i\cdot}$ by plugging in $\hat\delta_{ij}^0$ in the second equation of (3.20).

Our LRT statistic is then given as $\Lambda_*^{(12)}=-2\ln\Lambda^{(12)},$ where 
\begin{equation}
\Lambda^{(12)}=\frac{L(\hat\delta_{ij}^0,\hat\mu_{i\cdot}^0\mid \textrm { data})}{L(\hat\delta_{ij},\hat\beta_{ij}\mid \textrm { data})}=\frac{L(\hat\delta_{ij}^0,\hat\mu_{i\cdot}\mid X_{ijk} \hspace{0.1cm}\forall \hspace{0.1cm} i,j,k)}{L(\hat\delta_{ij},\hat\beta_{ij}\mid X_{ijk}\hspace{0.1cm}\forall\hspace{0.1cm} i,j,k)}.
\end{equation}
For `moderately large' $n_{ij}$ values, $\Lambda_*^{(12)}$ follows $\chi^2_{\nu_{12}}$ under $H_0^{(12)}$, with $\nu_{12}=a(b-1).$ The following Table 3.2 shows the simulated size of ALRT along with that of PBLRT (the proposed parametric bootstrap version based on $\Lambda_*^{(12)}$) for various combinations of parameters. (The steps of the PBLRT are similar to those under Case-1 with some obvious adjustments.)

\begin{table}[H] \caption{Simulated size of two tests with $a=b=2,\delta_{1\cdot}= \beta = 1.0$ (Problem - 1, Case-2)} 	
\fontsize{9pt}{12pt} 
	
	\begin{tabular}{|*{12}{c|}} \hline 
		\multirow{2}{*}{$\alpha$}&\multirow{2}{*}{$n_{ij}$}&\multicolumn{2}{c|}{$\delta_{2\cdot}=0.5$}&\multicolumn{2}{c|}{$\delta_{2\cdot}=1.0$}&\multicolumn{2}{c|}{$\delta_{2\cdot}=2.0$}&\multicolumn{2}{c|}{$\delta_{2\cdot}=5.0$}&\multicolumn{2}{c|}{$\delta_{2\cdot}=10.0$} \\ \cline{3-12}
		
		&&\multicolumn{1}{c|}{$\alpha_{ALRT}^{(12)}$}&$\alpha_{PBLRT}^{(12)}$&$\alpha_{ALRT}^{(12)}$&$\alpha_{PBLRT}^{(12)}$&$\alpha_{ALRT}^{(12)}$&$\alpha_{PBLRT}^{(12)}$&$\alpha_{ALRT}^{(12)}$&$\alpha_{PBLRT}^{(12)}$&$\alpha_{ALRT}^{(12)}$&$\alpha_{PBLRT}^{(12)}$ \\ \hline
		
		\multirow{4}{*}{$0.01$}
		
	&	5	&	0.025	&	0.008	&	0.028	&	0.010	&	0.026	&	0.010	&	0.025	&	0.010	&	0.021	&	0.008\\ \cline{2-12}
	&	10	&	0.020	&	0.011	&	0.017	&	0.010	&	0.018	&	0.012	&	0.016	&	0.010	&	0.017	&	0.011\\ \cline{2-12}
	&	25	&	0.015	&	0.011	&	0.016	&	0.012	&	0.015	&	0.012	&	0.013	&	0.010	&	0.012	&	0.009\\ \cline{2-12}
	&	50	&	0.012	&	0.012	&	0.011	&	0.010	&	0.011	&	0.012	&	0.013	&	0.011	&	0.012	&	0.010\\ \hline
		
		\multirow{4}{*}{$0.05$}
	&	5	&	0.100	&	0.045	&	0.105	&	0.054	&	0.097	&	0.050	&	0.095	&	0.049	&	0.088	&	0.047\\ \cline{2-12}
	&	10	&	0.073	&	0.050	&	0.070	&	0.048	&	0.078	&	0.055	&	0.074	&	0.048	&	0.066	&	0.046\\ \cline{2-12}
	&	25	&	0.059	&	0.051	&	0.065	&	0.055	&	0.058	&	0.050	&	0.058	&	0.048	&	0.057	&	0.049\\ \cline{2-12}
	&	50	&	0.059	&	0.056	&	0.057	&	0.053	&	0.058	&	0.053	&	0.058	&	0.054	&	0.058	&	0.054\\ \hline
		
		\multirow{4}{*}{$0.10$}
	&	5	&	0.168	&	0.095	&	0.176	&	0.101	&	0.170	&	0.100	&	0.171	&	0.099	&	0.161	&	0.094\\ \cline{2-12}
	&	10	&	0.131	&	0.095	&	0.126	&	0.093	&	0.139	&	0.106	&	0.139	&	0.103	&	0.128	&	0.096\\ \cline{2-12}
	&	25	&	0.111	&	0.096	&	0.117	&	0.105	&	0.116	&	0.102	&	0.108	&	0.096	&	0.103	&	0.094\\ \cline{2-12}
	&	50	&	0.108	&	0.102	&	0.111	&	0.104	&	0.113	&	0.106	&	0.108	&	0.101	&	0.106	&	0.100\\ \hline		
	\end{tabular}
\end{table}

\noindent{\bf{Remark 3.1}} For both the cases of Problem 1, the proposed PBLRT is showing a remarkable performance in maintaining the nominal level very closely, even for sample sizes as low as 5. On the other hand, the classical ALRT is performing poorly by being very liberal for small sample sizes (less than 25). For moderate to large 
sample sizes, both the tests have almost identical size values.

\section{\fontsize{13.2pt}{14pt}\selectfont Testing the joint significance of Factor - A and Factor - B (Problem - 3)}
\underline{\bfseries{Case-1}}: $\beta_{ij}=\beta \hspace{0.1cm}$ for all $\hspace{0.1cm} i,j.$

Our goal in this section is to consider the Problem - 3, with testing the null hypothesis $H_0^{(3)}:\mu_{ij}=\mu_{\cdot\cdot},\hspace{0.1cm}\forall\hspace{0.1cm}(i,j)$ against the alternative which negates it. The null hypothesis is stating that both the factors have no influence on the mean response when they act simultaneously, which under the assumption of equality of scales $(\beta_{ij}=\beta,\hspace{0.1cm}\forall\hspace{0.1cm}(i,j))$ can be written as
\begin{equation}
H_0^{(31)}:\mu_{ij}=\delta_{ij}\beta=\mu_{\cdot\cdot}\hspace{0.1cm}\forall(i,j) \textrm{ i.e., }    \delta_{ij}=\mu_{\cdot\cdot}/\beta=\delta_{\cdot\cdot}\textrm{ (say) for some suitable } \delta_{\cdot\cdot}.
\end{equation}
Similar to Section 2, the test statistic is given as
\begin{equation}
\Lambda_*^{(31)}=-2ln\{\sup_{H_0^{(31)}} {L}\}\big/\{\sup{L}\}.
\end{equation}
With all $n_{ij}$ `moderately large', we can approximate the sampling distribution of $\Lambda_*^{(31)}$ under $H_0^{(31)}$ as $\chi^2_{\nu 31}$ where the degrees of freedom $\nu_{31} = (ab-1).$

Next we are going to see the details of the LRT method with Problem - 3. The log-likelihood function under $H_0^{(31)},$ denoted by $L_*^{0(31)},$ is
\begin{equation}
L_*^{0(31)}=\sum_{i=1}^a\sum_{j=1}^b n_{ij}\{-ln\Gamma(\delta_{\cdot\cdot})-\delta_{\cdot\cdot}ln\beta-(1/\beta)\bar X_{ij\cdot}+(\delta_{\cdot\cdot}-1)ln\widetilde X_{ij\cdot}\}.
\end{equation}
By differentiating $L_*^{0(31)}$ w.r.t. $\delta_{\cdot\cdot}$ and $\beta,$ and then setting them equal to zero yields
\begin{equation}
\sum_{i=1}^a\sum_{j=1}^b n_{ij}\psi(\delta_{\cdot\cdot})=\sum_{i=1}^a\sum_{j=1}^b n_{ij}(ln\widetilde X_{ij\cdot}-ln\beta), \text{and}
\end{equation}
\begin{equation}
\beta\delta_{\cdot\cdot}\sum_{i=1}^a\sum_{j=1}^b n_{ij}=\sum_{i=1}^a\sum_{j=1}^b n_{ij}\bar X_{ij\cdot}
\end{equation}
The MLEs of $\delta_{\cdot\cdot}$ and $\beta$ under $H_0^{(31)},$ denoted by $\hat\delta_{\cdot\cdot}^0$ and $\hat\beta^0,$ are obtained as follows. First obtain $\hat\delta_{\cdot\cdot}$ by solving the following equation
\begin{equation}
\psi(\hat\delta_{\cdot\cdot}^0)-ln(\hat\delta_{\cdot\cdot}^0\sum_{i=1}^a\sum_{j=1}^b n_{ij})=\{(\sum_{i=1}^a\sum_{j=1}^b n_{ij}ln\widetilde X_{ij\cdot})\}\big/\{(\sum_{i=1}^a\sum_{j=1}^b n_{ij})\}-ln(\sum_{i=1}^a\sum_{j=1}^b n_{ij}\bar X_{ij\cdot})
\end{equation}
and then obtain $\hat \beta^0$ as
\begin{equation}
\hat{\beta}^0=(\sum_{i=1}^a\sum_{j=1}^b n_{ij}\bar X_{ij\cdot})\big/ (\delta_{\cdot\cdot}\sum_{i=1}^a\sum_{j=1}^b n_{ij}).
\end{equation}
Thus,
\begin{equation}
 \sup_{H_0^{(31)}}L=L(\hat \delta_{\cdot\cdot}^0,\hat \beta^0|X_{ijk}\hspace{0.1cm}\forall\hspace{0.1cm}(i,j,k))
\end{equation}
On the other hand,
\begin{equation}
\sup L=L(\hat \delta_{ij},\hat \beta|X_{ijk}\hspace{0.1cm}\forall\hspace{0.1cm}(i,j,k))
\end{equation}
Therefore, the LRT statistic is 
$\Lambda_*^{(31)}=2\ln\Lambda^{(31)},$ where
\begin{equation}
\Lambda^{(31)}=\frac{L(\hat \delta_{\cdot\cdot}^0,\hat \beta^0|X_{ijk}\hspace{0.1cm}\forall\hspace{0.1cm}(i,j,k))}{L(\hat \delta_{ij},\hat \beta|X_{ijk}\hspace{0.1cm}\forall\hspace{0.1cm}(i,j,k))}.
\end{equation}
The following Table 4.1 shows the size values of ALRT and PBLRT for various sample sizes.
 
\begin{table}[H] 
	\caption{Simulated size of two tests with $ a = b = 2;  \delta_{ij}=\delta_{\cdot\cdot}=1.0 =\beta$ (Problem - 3, Case-1)}
\fontsize{10pt}{12pt}
	\centering
			\begin{tabular}{|*{7}{c|}} \hline  
					\multirow{2}{*}{\diagbox[width=1.5cm]{$n_{ij}$}{$\alpha$}}
					&\multicolumn{2}{c|}{0.01}
					&\multicolumn{2}{c|}{0.05}
					&\multicolumn{2}{c|}{0.10} \\ \cline{2-7}
					&\multicolumn{1}{c|}{\quad \hspace{0cm} $\alpha_{ALRT}^{(31)}$ \hspace{0cm} \quad} & {\quad \hspace{0cm} $\alpha_{PBLRT}^{(31)}$\hspace{0cm} \quad} & {\quad \hspace{0cm} $\alpha_{ALRT}^{(31)}$ \hspace{0cm} \quad} & {\quad \hspace{0cm} $\alpha_{PBLRT}^{(31)}$\hspace{0cm} \quad} & {\quad \hspace{0cm} $\alpha_{ALRT}^{(31)}$ \hspace{0cm} \quad} & {\quad \hspace{0cm} $\alpha_{PBLRT}^{(31)}$\hspace{0cm} \quad}\\ \hline
					
					{5}&0.023 &0.011 & 0.091 & 0.051 &0.161 &0.100 \\ \hline
					{10}&0.017 &0.012 & 0.072 & 0.054  &0.131 & 0.107\\ \hline
					{25}&0.011 & 0.009 & 0.058 & 0.050  &0.111 &0.102 \\ \hline 
					{50}&0.013 &0.012 & 0.055 & 0.053 &0.110 &0.104 \\ \hline
				\end{tabular}
	
		\end{table}	
\underline{\bfseries{Case-2}}: ($\beta_{ij}$'s are not equal)

The log-likelihood function under $H_0^{(32)}:\delta_{ij}=\mu_{\cdot\cdot}/\beta_{ij},1\leq i \leq a,1\leq j \leq b$, henceforth denoted by $L_*^{0(32)}$ is
\begin{align}
	L_*^{0(32)} = \sum_{i=1}^a\sum_{j=1}^b n_{ij}\bigg\{- \ln\Gamma(\delta_{ij})-\delta_{ij}\ln\beta_{ij}-(1/\beta_{ij})\bar X_{ij\cdot}+(\delta_{ij}-1)\ln\tilde X_{ij\cdot} \bigg\}.
\end{align}
Differentiating $L_*^{0(32)}$ w.r.t $\delta_{ij}$ and $\mu_{\cdot\cdot}$, and then setting them equal to zero yields the estimates of $\delta_{ij}$ and $\mu_{\cdot\cdot}$ under $H_0^{(32)},$ denoted by $\hat\delta_{ij}^0$ and $\hat\mu_{\cdot\cdot}$
\begin{equation}
	\begin{aligned}
		\psi(\hat\delta_{ij}^0)
		&+ln\{
		(
		\sum_{i_0=1}^a\sum_{j_0=1}^b
		n_{i_0j_0}\hat\delta_{i_0j_0}^0\bar X_{i_0j_0\cdot}
		)
		\big/
		(
		\sum_{i_0=1}^a\sum_{j_0=1}^b
		n_{i_0j_0}\hat\delta_{i_0j_0}^0
		)
		\}
		\\
		&\quad+
		\bar X_{ij\cdot}
		\{
		(
		\sum_{i_0=1}^a\sum_{j_0=1}^b
		n_{i_0j_0}\hat\delta_{i_0j_0}^0
	)
		\big/
		(
		\sum_{i_0=1}^a\sum_{j_0=1}^b
		n_{i_0j_0}\hat\delta_{i_0j_0}^0\bar X_{i_0j_0\cdot}
		)
		\}
		=1+ln(\hat\delta_{ij}^0\tilde X_{ij\cdot}).
	\end{aligned}
\end{equation}
\begin{equation}
	 \textrm{ and } \hat\mu_{\cdot\cdot}=\{(\sum_{i=1}^a\sum_{j=1}^b n_{ij} \hat\delta_{ij}^0\bar X_{ij\cdot})\big/(\sum_{i=1}^a\sum_{j=1}^bn_{ij}\hat\delta_{ij}^0)\}.
\end{equation}

\begin{table}[H] 
	\caption{Simulated size of two tests with $ a = b = 2;  \delta_{ij}=\delta_{\cdot\cdot}=1.0 =\beta$ (Problem - 3, Case-2)}
\fontsize{10pt}{12pt}
	\centering
		\begin{tabular}{|*{7}{c|}} \hline  
			\multirow{2}{*}{\diagbox[width=1.5cm]{$n_{ij}$}{$\alpha$}}
			&\multicolumn{2}{c|}{0.01}
			&\multicolumn{2}{c|}{0.05}
			&\multicolumn{2}{c|}{0.10} \\ \cline{2-7}
			&\multicolumn{1}{c|}{\quad \hspace{0cm} $\alpha_{ALRT}^{(32)}$ \hspace{0cm} \quad} & {\quad \hspace{0cm} $\alpha_{PBLRT}^{(32)}$\hspace{0cm} \quad} & {\quad \hspace{0cm} $\alpha_{ALRT}^{(32)}$ \hspace{0cm} \quad} & {\quad \hspace{0cm} $\alpha_{PBLRT}^{(32)}$\hspace{0cm} \quad} & {\quad \hspace{0cm} $\alpha_{ALRT}^{(32)}$ \hspace{0cm} \quad} & {\quad \hspace{0cm} $\alpha_{PBLRT}^{(32)}$\hspace{0cm} \quad}\\ \hline
			
			{5}&0.110&0.009&0.206&0.047&0.285&0.091 \\ \hline
			{10}&0.045&0.009&0.114 &0.052&0.176&0.098\\ \hline
			{25}&0.022&0.010&0.078&0.057&0.135&0.105 \\ \hline 
			{50}&0.014&0.010&0.056&0.047& 0.110 &0.097 \\ \hline
		\end{tabular}

\end{table}
\noindent The ALRT rejects $H_0$ if $\Lambda^{(31)}>\chi^2_{\nu_{32},\alpha}$, with $\nu_{32}=(ab-1)$.	\\

\noindent\textbf {Remark 4.1} Once again, the trends observed for Problem - 3 are almost similar to those seen for Problem - 1 (recall Remark 3.1). Extensive simulations have been carried out for various combinations of the parameter values, but only limited results are being shown here for brevity.
\section{\fontsize{14.5pt}{14pt}\selectfont Testing the interaction of two factors in multiplicative form (Problem - 4)}
In this section, we will consider the Problem - 4: $H_0^{(4)}: \mu_{ij} =\mu_{i\cdot}\times\mu_{\cdot j} $ against the alternative which negates it, under both the subcases, when $\beta_{ij}\textrm{'}s$ are all equal and when $\beta_{ij}\textrm{'}s$ are not equal.

\underline{\bfseries{Case-1}}: $\beta_{ij}=\beta$ for all $i,j$.\\
In this case, the null hypothesis can be written as $H_0^{(41)}:\delta_{ij}=\delta_{i\cdot}\times\delta_{\cdot j}$. So, under $H_0^{(41)}$, the log-likelihood function as follows
\begin{align}
L_*^{0(41)} = \sum_{i=1}^a\sum_{j=1}^b n_{ij}\bigg\{- ln\Gamma(\delta_{i\cdot}\times\delta_{\cdot j})-(\delta_{i\cdot}\times\delta_{\cdot j})ln\beta-1/\beta\bar X_{ij\cdot}+(\delta_{i\cdot}\times\delta_{\cdot j}-1)ln\tilde X_{ij\cdot} \bigg\}.
\end{align}
By differentiating $L_*^{0(41)}$ in (5.1) w.r.t $\hat{\delta}_{i\cdot}^0,\hat{\delta}_{\cdot j}^0,$ and $\hat{\beta}^0$ and then setting them equal to zero yields the following system of equations
\begin{align}
	\sum_{j=1}^b n_{ij}\hat{\delta}_{\cdot j}^0
	\psi(\hat{\delta}_{i\cdot}^0\times\hat{\delta}_{\cdot j}^0)
	&=
	\sum_{j=1}^b n_{ij}\hat{\delta}_{\cdot j}^0
	ln\bigl\{
	\bigl(
	\tilde X_{ij\cdot}
	\sum_{i_0=1}^a\sum_{j_0=1}^b
	n_{i_0j_0}
	(\hat{\delta}_{i_0\cdot}^0\times\hat{\delta}_{\cdot j_0}^0)
	\bigr)
	\big/
	\nonumber\\
	&\qquad
	\bigl(
	\sum_{i_0=1}^a\sum_{j_0=1}^b
	n_{i_0j_0}\bar X_{i_0j_0\cdot}
	\bigr)
	\bigr\},
	\qquad i=1,2,\ldots,a.
\end{align}
\begin{align}
	\sum_{i=1}^a n_{ij}\hat{\delta}_{i\cdot}^0
	\psi(\hat{\delta}_{i\cdot}^0\times\hat{\delta}_{\cdot j}^0)
	&=
	\sum_{i=1}^a n_{ij}\hat{\delta}_{i\cdot}^0
	ln\{
	\bigl(
	\tilde X_{ij\cdot}
	\sum_{i_0=1}^a\sum_{j_0=1}^b
	n_{i_0j_0}
	(\hat{\delta}_{i_0\cdot}^0 \times \hat{\delta}_{\cdot j_0}^0
	)\bigr)
	\big/
	\nonumber\\
	&\qquad\qquad
	\bigr(
	\sum_{i_0=1}^a\sum_{j_0=1}^b
	n_{i_0j_0}\bar X_{i_0j_0\cdot}
	\bigr)
	\},
	\qquad \forall\, j=1,2,\ldots,b.
\end{align}

\begin{align}
\hat{\beta}^0=\{(\sum_{i=1}^a\sum_{j=1}^b n_{ij}\bar X_{ij\cdot})\big/ (\sum_{i=1}^a\sum_{j=1}^b n_{ij}(\hat{\delta}_{i\cdot}^0\times\hat{\delta}_{\cdot j}^0)\}
\end{align}
The degrees of freedom is $\nu=(ab+1)-(a+b+1)=ab-a-b=(a-1)(b-1)-1$, provided $max\{a,b\}>2$.\\
Thus,
\begin{equation}
	\sup_{H_0^{(41)}}L=L(\hat \delta_{i\cdot}^0,\hat \delta_{\cdot j}^0,\hat \beta^0|X_{ijk}\hspace{0.1cm}\forall\hspace{0.1cm}(i,j,k))
\end{equation}
On the other hand,
\begin{equation}
	\sup L=L(\hat \delta_{ij},\hat \beta|X_{ijk}\hspace{0.1cm}\forall\hspace{0.1cm}(i,j,k))
\end{equation}
Then write the LRT statistics $\Lambda_*^{(41)}=2\ln\Lambda^{(41)},$ where
\begin{equation}
	\Lambda^{(41)}=\frac{L(\hat \delta_{i\cdot}^0,\hat \delta_{\cdot j}^0,\hat \beta^0|X_{ijk}\hspace{0.1cm}\forall\hspace{0.1cm}(i,j,k))}{L(\hat \delta_{ij},\hat \beta|X_{ijk}\hspace{0.1cm}\forall\hspace{0.1cm}(i,j,k))}.
\end{equation}

\begin{table*}[ht] 
	\caption{Simulated size of two tests with $a=3, b=2,\delta_{1\cdot}=1.0, \delta_{2\cdot}=0.5,\delta_{3\cdot}=2.0, \delta_{\cdot1}=1.0,\delta_{\cdot2}=2.0, \beta = 1.0$ (Problem - 4, Case-1)} 
	\fontsize{10pt}{12pt}
	\centering
		\begin{tabular}{|*{7}{c|}} \hline  
			\multirow{2}{*}{\diagbox[width=1.5cm]{$n_{ij}$}{$\alpha$}}
			&\multicolumn{2}{c|}{0.01}
			&\multicolumn{2}{c|}{0.05}
			&\multicolumn{2}{c|}{0.10} \\ \cline{2-7}
			&\multicolumn{1}{c|}{\quad \hspace{0cm} $\alpha_{ALRT}^{(41)}$ \hspace{0cm} \quad} & {\quad \hspace{0cm} $\alpha_{PBLRT}^{(41)}$\hspace{0cm} \quad} & {\quad \hspace{0cm} $\alpha_{ALRT}^{(41)}$ \hspace{0cm} \quad} & {\quad \hspace{0cm} $\alpha_{PBLRT}^{(41)}$\hspace{0cm} \quad} & {\quad \hspace{0cm} $\alpha_{ALRT}^{(41)}$ \hspace{0cm} \quad} & {\quad \hspace{0cm} $\alpha_{PBLRT}^{(41)}$\hspace{0cm} \quad}\\ \hline
			
			{5} &0.069  &0.010	&	0.215	&0.055	&0.338	&0.108 \\ \hline
			{10}&0.048	&0.010	&0.183	&	0.051	&	0.297	&0.107 \\ \hline
			{25}&0.039	&0.012	&0.159	&0.055	&0.273	&0.100 \\ \hline
			{50}&0.045	&0.011	&0.146	&0.046	&0.275	&0.106 \\ \hline
			
		\end{tabular}

\end{table*}

\underline{\bfseries{Case-2}:} $\beta_{ij}$\textrm{'}s are not equal\\
The log-likelihood function under $H_0^{(42)}:$
\begin{align}
	L_*^{0(42)}
	&=
	\sum_{i=1}^a\sum_{j=1}^b n_{ij}
	\bigl\{
	-ln\Gamma\Bigl((\mu_{i\cdot}\times\mu_{\cdot j})/\beta_{ij}\Bigr)
	-\bigl((\mu_{i\cdot}\times\mu_{\cdot j})/\beta_{ij}\bigr)ln\beta_{ij}
	\nonumber\\
	&\qquad
	-\bar X_{ij\cdot}/\beta_{ij}
	+\Bigl((\mu_{i\cdot}\times\mu_{\cdot j})/\beta_{ij}-1\Bigr)
	ln\tilde X_{ij\cdot}
	\bigr\}.
\end{align}

By differentiating $L_*^{0(42)}$ in (5.5) w.r.t $\hat{\mu}_{i\cdot}^0,\hat{\mu}_{\cdot j}^0,$ and $\hat{\beta}_{ij}^0$ and then setting them equal to zero yields the following system of equations
\begin{align}
\sum_{j=1}^b n_{ij}(\hat{\mu}_{\cdot j}^0/\hat{\beta}_{ij}^0)\psi\left((\hat{\mu}_{i\cdot}^0\times\hat{\mu}_{\cdot j}^0)/\hat{\beta}_{ij}^0\right)=\sum_{j=1}^b n_{ij}(\hat{\mu}_{\cdot j}^0/\hat{\beta}_{ij}^0)\left(ln\tilde X - ln\hat{\beta}_{ij}^0\right)
\end{align}
\begin{align}
\sum_{i=1}^a n_{ij}(\hat{\mu}_{i\cdot }^0/\hat{\beta}_{ij}^0)\psi\left((\hat{\mu}_{i\cdot}^0\times\hat{\mu}_{\cdot j}^0)/\hat{\beta}_{ij}^0\right)=\sum_{i=1}^a n_{ij}(\hat{\mu}_{i\cdot }^0/\hat{\beta}_{ij}^0)\left(ln\tilde X - ln\hat{\beta}_{ij}^0\right)
\end{align}
\begin{align}
\psi\left((\hat{\mu}_{i\cdot}^0\times\hat{\mu}_{\cdot j}^0)/\hat{\beta}_{ij}^0\right)+ln(\hat{\beta}_{ij}^0/\tilde X_{ij\cdot})-1+\{\bar X_{ij\cdot}/(\hat{\mu}_{i\cdot}^0\times\hat{\mu}_{\cdot j}^0)\}=0
\end{align}
The degree of freedom is $\nu=2ab-(a+b+ab)=ab-a-b$, provided $max\{a,b\}>2$.\\
Thus,
\begin{equation}
	\sup_{H_0^{(42)}}L=L(\hat \delta_{i\cdot}^0,\hat \delta_{\cdot j}^0,\hat \beta_{ij}^0|X_{ijk}\hspace{0.1cm}\forall\hspace{0.1cm}(i,j,k))
\end{equation}
On the other hand,
\begin{equation}
	\sup L=L(\hat \delta_{ij},\hat \beta|X_{ijk}\hspace{0.1cm}\forall\hspace{0.1cm}(i,j,k))
\end{equation}
Then write the LRT statistics $\Lambda_*^{(42)}=2\ln\Lambda^{(42)},$ where
\begin{equation}
	\Lambda^{(42)}=\frac{L(\hat \mu_{i\cdot}^0,\hat \mu_{\cdot j}^0,\hat \beta_{ij}^0|X_{ijk}\hspace{0.1cm}\forall\hspace{0.1cm}(i,j,k))}{L(\hat \delta_{ij},\hat \beta_{ij}|X_{ijk}\hspace{0.1cm}\forall\hspace{0.1cm}(i,j,k))}.
\end{equation}

\begin{table}[H] \caption{Simulated size of two tests with $a=3, b=2,\delta_{1\cdot}=4.0, \delta_{2\cdot}=5.0,\delta_{3\cdot}=6.0, \delta_{\cdot1}=3.0,\delta_{\cdot2}=4.0, \beta = 1.0$ (Problem - 4, Case-2)} 
	\fontsize{10pt}{12pt}
	\centering
		\begin{tabular}{|*{7}{c|}} \hline  
			\multirow{2}{*}{\diagbox[width=1.5cm]{$n_{ij}$}{$\alpha$}}
			&\multicolumn{2}{c|}{0.01}
			&\multicolumn{2}{c|}{0.05}
			&\multicolumn{2}{c|}{0.10} \\ \cline{2-7}
			&\multicolumn{1}{c|}{\quad \hspace{0cm} $\alpha_{ALRT}^{(42)}$ \hspace{0cm} \quad} & {\quad \hspace{0cm} $\alpha_{PBLRT}^{(42)}$\hspace{0cm} \quad} & {\quad \hspace{0cm} $\alpha_{ALRT}^{(42)}$ \hspace{0cm} \quad} & {\quad \hspace{0cm} $\alpha_{PBLRT}^{(42)}$\hspace{0cm} \quad} & {\quad \hspace{0cm} $\alpha_{ALRT}^{(42)}$ \hspace{0cm} \quad} & {\quad \hspace{0cm} $\alpha_{PBLRT}^{(42)}$\hspace{0cm} \quad}\\ \hline
			
			{5}&0.065	&0.007	&0.216	&0.045	&0.345	&0.091 \\ \hline
			{10}&0.048	&0.010	&0.184	&0.047	&0.302	&0.101 \\ \hline
			{25}&0.040	&0.009	&0.161	&0.050	&0.283	&0.101 \\ \hline
			{50}&0.040	&0.010	&0.154	&0.050	&0.273	&0.099 \\ \hline
		\end{tabular}

\end{table}
\noindent\textbf{Remark 5.1} For both cases considered in Problem - 4, the proposed PBLRT demonstrates remarkable performance in maintaining the nominal significance level, even for very small sample sizes as low as 5. In contrast, the classical ALRT performs poorly, exhibiting a highly liberal behavior when the sample size is small (less than 25). As the sample size increases, the significance level of the ALRT gradually converges to the nominal level $\alpha$, albeit slowly.

\section{Revisiting the Eight Examples Using ANOGaM-2}
We now revisit the eight examples discussed in Section 1 and have them analyzed using the ANOGaM-2 technique. Not only first we compare the decisions based on ALRT vs. PBLRT, but also see the similarities and dissimilarities with those under the usual normality-based ANOVA results.

The following Table 6.1 summarizes the results in terms of $P$-values based on ALRT as well as PBLRT under the proposed gamma model. For each example (or dataset) we use the notation $\Lambda_*^{(k\ell)}$ to denote the LRT statistic with the superscript ``$(k\ell)$", where $k$ indicates the `problem number' ($1, 2, 3\text{ or } 4$), and $\ell$ indicates the 'case number' ($1 = \text{scale parameters are all equal}$, $2 = \text{scale parameters are not equal}$). (In the given context, $k=3$ was not encountered.) The value of $\ell$ is determined by Problem - 5 which, as shown in Section 2, indicated (based on the corresponding PBLRT) that only for Example 1.4 we need to use $\ell = 2$ (i.e., unequal scale parameters). Table 6.1 should be seen in conjunction with Table 2.2.
\small
\setlength{\tabcolsep}{2.5pt}

\begin{longtable}{|c|c||c|c||c|c|}
	\caption{$P$-values for eight examples using Gamma ALRT and PBLRT}
	\label{tab:pvalues-gamma}\\
	\hline
	\textbf{Example}
	& $\boldsymbol{\Lambda_*^{(k\ell)}}$
	& $\boldsymbol{P_{ALRT}}$
	& \makecell{\textbf{Decision based}\\\textbf{on ALRT}}
	& $\boldsymbol{P_{PBLRT}}$
	& \makecell{\textbf{Decision based}\\\textbf{on PBLRT}}\\
	[-2pt]\hline
	\endfirsthead
	
	\multicolumn{6}{c}%
	{{\tablename\ \thetable{} -- continued from previous page}}\\
	\hline
	\textbf{Example}
	& $\boldsymbol{\Lambda_*^{(i1)}}$
	& $\boldsymbol{P_{ALRT}}$
	& \makecell{\textbf{Decision based}\\\textbf{on ALRT}}
	& $\boldsymbol{P_{PBLRT}}$
	& \makecell{\textbf{Decision based}\\\textbf{on PBLRT}}\\
	[-2pt] \hline
	\endhead
	
	\hline
	\multicolumn{6}{r}{{Continued on next page}}\\
	\endfoot
	
	\hline
	\endlastfoot
	
	\multirow{3}{*}{\makecell{1.1\\(Under Case-1\\of Problem - 5)}}
	& $\Lambda_*^{(11)}=39.484$
	& $P_{ALRT}^B<0.001$
	& \makecell{Problem - 1:\\Reject null}
	& $P_{PBLRT}^B<0.001$
	& \makecell{Problem - 1:\\Reject null}
	\\ \cline{2-6}
	
	& $\Lambda_*^{(21)}=133.291$
	& $P_{ALRT}^A<0.001$
	& \makecell{Problem - 2:\\Reject null}
	& $P_{PBLRT}^A<0.001$
	& \makecell{Problem - 2:\\Reject null}
	\\ \cline{2-6}
	
	& $\Lambda_*^{(41)}=1.909$
	& $P_{ALRT}^{AB}=0.752$
	& \makecell{Problem - 4:\\{\bf{Retain null}}}
	& $P_{PBLRT}^{AB}<0.001$
	& \makecell{Problem - 4:\\{\bf{Reject null}}}
	\\ \hline

	\multirow{3}{*}{\makecell{1.2\\(Under Case-1\\of Problem - 5)}}
	& $\Lambda_*^{(11)}=34.039$
	& $P_{ALRT}^B<0.001$
	& \makecell{Problem - 1:\\Reject null}
	& $P_{PBLRT}^B<0.001$
	& \makecell{Problem - 1:\\Reject null}
	\\ \cline{2-6}
	
	& $\Lambda_*^{(21)}=71.851$
	& $P_{ALRT}^A<0.001$
	& \makecell{Problem - 2:\\Reject null}
	& $P_{PBLRT}^A<0.001$
	& \makecell{Problem - 2:\\Reject null}
	\\ \cline{2-6}
	
	& $\Lambda_*^{(41)}=2.482$
	& $P_{ALRT}^{AB}=0.115$
	& \makecell{Problem - 4:\\{\bf{Retain null}}}
	& $P_{PBLRT}^{AB}<0.001$
	& \makecell{Problem - 4:\\{\bf{Reject null}}}
	\\ \hline

	\multirow{3}{*}{\makecell{1.3\\ (Under Case-1\\of Problem - 5)}}
	& $\Lambda_*^{(11)}=44.881$
	& $P_{ALRT}^B<0.001$
	& \makecell{Problem - 1:\\Reject null}
	& $P_{PBLRT}^B<0.001$
	& \makecell{Problem - 1:\\Reject null}
	\\ \cline{2-6}
	
	& $\Lambda_*^{(21)}=96.821$
	& $P_{ALRT}^A<0.001$
	& \makecell{Problem - 2:\\Reject null}
	& $P_{PBLRT}^A<0.001$
	& \makecell{Problem - 2:\\Reject null}
	\\ \cline{2-6}
	
	& $\Lambda_*^{(41)}=29.795$
	& $P_{ALRT}^{AB}<0.001$
	& \makecell{Problem - 4:\\Reject null}
	& $P_{PBLRT}^{AB}<0.001$
	& \makecell{Problem - 4:\\Reject null}
	\\ \hline

	\multirow{3}{*}{\makecell{1.4\\(Under Case-2\\of Problem - 5)}}
	& $\Lambda_*^{(12)}=49.203$
	& $P_{ALRT}^B<0.001$
	& \makecell{Problem - 1:\\Reject null}
	& $P_{PBLRT}^B<0.001$
	& \makecell{Problem - 1:\\Reject null}
	\\ \cline{2-6}
	
	& $\Lambda_*^{(22)}=31.866$
	& $P_{ALRT}^A<0.001$
	& \makecell{Problem - 2:\\Reject null}
	& $P_{PBLRT}^A<0.001$
	& \makecell{Problem - 2:\\Reject null}
	\\ \cline{2-6}
	
	& $\Lambda_*^{(42)}=19.751$
	& $P_{ALRT}^{AB}<0.001$
	& \makecell{Problem - 4:\\Reject null}
	& $P_{PBLRT}^{AB}<0.001$
	& \makecell{Problem - 4:\\Reject null}
	\\ \hline

	\multirow{3}{*}{\makecell{1.5\\(Under Case-1\\of Problem - 5)}}
	& $\Lambda_*^{(11)}=1.844$
	& $P_{ALRT}^B=0.764$
	& \makecell{Problem - 1:\\{\bf{Retain null}}}
	& $P_{PBLRT}^B<0.001$
	& \makecell{Problem - 1:\\{\bf{Reject null}}}
	\\ \cline{2-6}
	
	& $\Lambda_*^{(21)}=1.186$
	& $P_{ALRT}^A=0.756$
	& \makecell{Problem - 2:\\{\bf{Retain null}}}
	& $P_{PBLRT}^A<0.001$
	& \makecell{Problem - 2:\\{\bf{Reject null}}}
	\\ \cline{2-6}
	
	& $\Lambda_*^{(41)}=0.722$
	& $P_{ALRT}^{AB}=0.396$
	& \makecell{Problem - 4:\\{\bf{Retain null}}}
	& $P_{PBLRT}^{AB}<0.001$
	& \makecell{Problem - 4:\\{\bf{Reject null}}}
	\\ \hline

	\multirow{3}{*}{\makecell{1.6\\(Under Case-1\\of Problem - 5)}}
	& $\Lambda_*^{(11)}=64.658$
	& $P_{ALRT}^B<0.001$
	& \makecell{Problem - 1:\\Reject null}
	& $P_{PBLRT}^B<0.001$
	& \makecell{Problem - 1:\\Reject null}
	\\ \cline{2-6}
	
	& $\Lambda_*^{(21)}=34.711$
	& $P_{ALRT}^A<0.001$
	& \makecell{Problem - 2:\\Reject null}
	& $P_{PBLRT}^A<0.001$
	& \makecell{Problem - 2:\\Reject null}
	\\ \cline{2-6}
	
	& $\Lambda_*^{(41)}=33.865$
	& $P_{ALRT}^{AB}<0.001$
	& \makecell{Problem - 4:\\Reject null}
	& $P_{PBLRT}^{AB}<0.001$
	& \makecell{Problem - 4:\\Reject null}
	\\ \hline

	\multirow{3}{*}{\makecell{1.7\\(Under Case-1\\of Problem - 5)}}
	& $\Lambda_*^{(11)}=51.108$
	& $P_{ALRT}^B<0.001$
	& \makecell{Problem - 1:\\Reject null}
	& $P_{PBLRT}^B<0.001$
	& \makecell{Problem - 1:\\Reject null}
	\\ \cline{2-6}
	
	& $\Lambda_*^{(21)}=33.794$
	& $P_{ALRT}^A<0.001$
	& \makecell{Problem - 2:\\Reject null}
	& $P_{PBLRT}^A<0.001$
	& \makecell{Problem - 2:\\Reject null}
	\\ \cline{2-6}
	
	& $\Lambda_*^{(41)}=25.864$
	& $P_{ALRT}^{AB}<0.001$
	& \makecell{Problem - 4:\\Reject null}
	& $P_{PBLRT}^{AB}<0.001$
	& \makecell{Problem - 4:\\Reject null}
	\\ \hline
	
	\multirow{3}{*}{\makecell{1.8\\(Under Case-1\\of Problem - 5)}}
	& $\Lambda_*^{(11)}=23.195$
	& $P_{ALRT}^B=0.003$
	& \makecell{Problem - 1:\\Reject null}
	& $P_{PBLRT}^B<0.001$
	& \makecell{Problem - 1:\\Reject null}
	\\ \cline{2-6}
	
	& $\Lambda_*^{(21)}=79.849$
	& $P_{ALRT}^A<0.001$
	& \makecell{Problem - 2:\\Reject null}
	& $P_{PBLRT}^A<0.001$
	& \makecell{Problem - 2:\\Reject null}
	\\ \cline{2-6}
	
	& $\Lambda_*^{(41)}=12.164$
	& $P_{ALRT}^{AB}=0.033$
	& \makecell{Problem - 4:\\Reject null}
	& $P_{PBLRT}^{AB}<0.001$
	& \makecell{Problem - 4:\\Reject null}
	\\	
\end{longtable}

As discussed earlier in Section 1, the purpose of this work is to present a different way of analyzing nonnegative observations subject to two factors using the gamma model especially when the sample sizes are "small". However, the proposed ANOGaM-2 method may or may not agree with the usual ANOVA approach in terms of drawing the final inferences. The following Table~\ref{tab:gamma_normal_comparison} summarizes these outcomes based on both the above methods. Note that under the gamma model PBLRT should be used as it maintains the level condition more accurately. The ALRT decisions based on the gamma model have been shown (in the middle column under each effect) just for completeness.\\

\begin{sidewaystable}[htbp]
	\centering
	\caption{Comparison of the outcomes : 2-Factor ANOVA vs. ANOGaM-2}
	\label{tab:gamma_normal_comparison}
	\renewcommand{\arraystretch}{1.3}
	\resizebox{\textheight}{!}{%
	\begin{tabular}{|c|c|c|c||c|c|c||c|c|c||}
		\hline
		\multirow{2}{*}{Data}
		& \multicolumn{3}{c||}{Factor-A Effect}
		& \multicolumn{3}{c||}{Factor-B Effect}
		& \multicolumn{3}{c||}{AB-Interaction Effect} \\
		\cline{2-10}
		& Normal & Gamma (ALRT) & Gamma (PBLRT)
		& Normal & Gamma (ALRT) & Gamma (PBLRT)
		& Normal & Gamma (ALRT) & Gamma (PBLRT) \\
		\hline
		Example 1.1
		& significant & significant & significant
		& significant & significant & significant
		& insignificant &insignificant & insignificant \\
		\hline
		Example 1.2
		& significant & significant & significant
		& significant & significant & significant
		& {\bf{significant}} &{\bf{insignificant}} &{\bf{insignificant}}  \\
		\hline
		Example 1.3
		& significant & significant & significant
		& significant & significant & significant
		& significant &significant &significant  \\
		\hline
		Example 1.4
		& significant & significant & significant
		& significant & significant & significant
		& significant &significant & significant\\
		\hline
		Example 1.5
		& {\bf{insignificant}} & {\bf{insignificant}} & {\bf{significant}}
		& {\bf{insignificant}} & {\bf{insignificant}} & {\bf{significant}}
		& insignificant &insignificant &insignificant \\
		\hline
		Example 1.6
		& {\bf{insignificant}} & {\bf{significant}} & {\bf{significant}}
		& significant & significant & significant
		& significant &significant &significant \\
		\hline
		Example 1.7
		& significant & significant & significant
		& significant & significant & significant
		& significant &significant &significant \\
		\hline
		Example 1.8
		& significant & significant & significant
		& significant & significant & significant
		&{\bf{insignificant}} &{\bf{significant}} & {\bf{significant}}\\		
		\hline
	\end{tabular}
}
\end{sidewaystable}

\noindent\textbf{Remark 6.1}
 In a nut-shell, the below Table~\ref{tab:gamma_normal_comparison} throws up an interesting scenario. Out of the eight examples, in four both the methods provide the same outcomes (see Examples 1.1, 1.3, 1.4 and 1.7). Recall that in all but Example 1.4 the gamma model used the equality of scale parameters. In spite of unequal scale parameters, ANOGaM-2 inferences agree with those of ANOVA in Example 1.4. For the remaining four examples (see Examples 1.2, 1.5, 1.6 and 1.8) ANOVA and ANOGaM-2 have varying levels of disagreements. Out of these four examples, Examples 1.5 and 1.6 appeared to have cleared the normality and homoscedasticity assumptions, whereas Examples 1.2 and 1.8 had witnessed doubts about these regular assumptions. Also, this study of four examples sends a powerful message that the standard two-way ANOVA should not be used blindly for nonnegative observations as a more robust (and/or flexible) gamma model provides an alternative avenue to analyze such data. Half of the eight examples have yielded disagreements between ANOVA and ANOGaM-2, and the applied researchers should take note of this.\\
 
 \noindent\textbf{Concluding Remark} This work provides an alternative tool, that is - ANOGaM-2, to the applied researchers dealing with nonnegative observations under two factors, thereby "thinking outside the box," i.e., going beyond the traditional homoscedastic normality based two-way ANOVA method. For a given dataset if both the methods yield the same inference(s) then such results would be more reliable. In case of disagreement(s) one should check carefully the potential causes behind such conflicting outcome(s). Nevertheless, the proposed ANOGaM-2 is more appealing as it is free from the restrictions under the ANOVA set-up, and tends to have more flexibility even though it is applicable only for nonnegative observations.
\newpage
\appendix
\renewcommand{\thesection}{Appendix~\Alph{section}}
\section{The codes of the simulation Problem - 5}

\begin{lstlisting}[style=Rcompact,
	caption={Simulation code for Table 2.1},
	label={lst:table21}]
	
	# ============================================================
	# Simulation for Table 2.1
	# ============================================================
	
	rm(list = ls(all = TRUE))
	library(nleqslv)
	
	myseed <- 123456
	set.seed(myseed)
	
	# Counters
	count  <- 0  # First call within each q
	count2 <- 0  # PBLRT rejections
	count3 <- 0  # ALRT rejections
	
	# Simulation settings
	a <- b <- 2
	n_ij <- 5
	alpha <- 0.01
	delta1 <- delta2 <- delta3 <- delta4 <- 2
	Q <- M <- 5000
	beta <- 1
	
	# Data and bootstrap parameter containers
	x <- array(NA_real_, c(a, b, n_ij))
	delta_in <- matrix(c(delta1, delta2, delta3, delta4), a, b)
	delta_gen <- matrix(NA_real_, a, b)
	beta_gen <- numeric(1)
	
	# ============================================================
	# Root-bracketing function
	# ============================================================
	
	bracket_root <- function(f, lower = .Machine$double.eps,
	upper = 5e4, max_expand = 8) {
		
		f_low <- suppressWarnings(
		tryCatch(f(lower), error = function(e) NA_real_)
		)
		f_up <- suppressWarnings(
		tryCatch(f(upper), error = function(e) NA_real_)
		)
		
		if (!is.na(f_low) && !is.na(f_up) && f_low * f_up <= 0)
		return(c(lower, upper))
		
		u <- upper
		for (k in 1:max_expand) {
			u <- u * 10
			f_up <- suppressWarnings(
			tryCatch(f(u), error = function(e) NA_real_)
			)
			if (!is.na(f_low) && !is.na(f_up) && f_low * f_up <= 0)
			return(c(lower, u))
		}
		
		l <- lower
		for (k in 1:max_expand) {
			l <- max(l / 10, .Machine$double.eps)
			f_low <- suppressWarnings(
			tryCatch(f(l), error = function(e) NA_real_)
			)
			if (!is.na(f_low) && !is.na(f_up) && f_low * f_up <= 0)
			return(c(l, u))
		}
		
		return(NA)
	}
	
	# ============================================================
	# Function to calculate lambda_star
	# ============================================================
	
	cal_lambda <- function(a, b, n_ij, x) {
		
		# 1. Arithmetic and geometric means
		x_bar <- matrix(NA_real_, a, b)
		x_tilde <- matrix(NA_real_, a, b)
		
		for (i in 1:a) for (j in 1:b) {
			xx <- x[i, j, ]
			x_bar[i, j] <- mean(xx, na.rm = TRUE)
			x_tilde[i, j] <- exp(mean(log(xx), na.rm = TRUE))
		}
		
		# 2. Estimate delta_ij and beta_ij from Equation (2.7)
		delta <- matrix(NA_real_, a, b)
		beta <- matrix(NA_real_, a, b)
		
		for (i in 1:a) for (j in 1:b) {
			f1 <- function(a_) {
				digamma(a_) - log(a_) -
				log(x_tilde[i, j] / x_bar[i, j])
			}
			
			br <- bracket_root(f1, lower = .Machine$double.eps,
			upper = 5e4)
			if (all(is.na(br))) return(NA_real_)
			
			root <- suppressWarnings(
			tryCatch(uniroot(f1, interval = br)$root,
			error = function(e) NA_real_)
			)
			if (is.na(root)) return(NA_real_)
			
			delta[i, j] <- root
			beta[i, j] <- x_bar[i, j] / delta[i, j]
		}
		
		# 3. Log-likelihood L1: Equation (3.16)
		L_star <- matrix(0, a, b)
		
		for (i in 1:a) for (j in 1:b) {
			L_star[i, j] <- n_ij * (
			-lgamma(delta[i, j]) -
			delta[i, j] * log(beta[i, j]) -
			x_bar[i, j] / beta[i, j] +
			(delta[i, j] - 1) * log(x_tilde[i, j])
			)
		}
		L1 <- sum(L_star)
		
		# 4. Solve Equation (2.8) for delta_0
		solve_system <- function(n_ij, x_bar, x_tilde) {
			
			i_size <- nrow(x_bar)
			j_size <- ncol(x_bar)
			
			equations <- function(delta_tam) {
				deltaM <- matrix(delta_tam, i_size, j_size)
				eqs <- matrix(NA_real_, i_size, j_size)
				
				for (ii in 1:i_size) for (jj in 1:j_size) {
					eqs[ii, jj] <- digamma(deltaM[ii, jj]) -
					log(sum(n_ij * deltaM)) -
					log(x_tilde[ii, jj] / sum(n_ij * x_bar))
				}
				as.vector(eqs)
			}
			
			initial_num <- runif(1, 0.001, 1)
			initial_guess <- rep(initial_num, i_size * j_size)
			
			sol <- try(
			nleqslv::nleqslv(
			initial_guess, equations,
			method = "Broyden",
			control = list(maxit = 2000, trace = 0)
			),
			silent = TRUE
			)
			
			if (inherits(sol, "try-error")) return(NA)
			if (is.null(sol$termcd) || sol$termcd != 1) return(NA)
			
			matrix(sol$x, i_size, j_size)
		}
		
		delta_0 <- solve_system(n_ij, x_bar, x_tilde)
		if (any(is.na(delta_0))) return(NA_real_)
		
		# 5. Calculate beta_0: Equation (2.9)
		beta_0 <- sum(n_ij * x_bar) / sum(n_ij * delta_0)
		
		# Store estimates from the first call within each q
		if (count == 0) {
			delta_gen <<- delta_0
			beta_gen <<- beta_0
			count <<- count + 1
		}
		
		# 6. Log-likelihood L2: Equation (3.19)
		L0_star <- matrix(NA_real_, a, b)
		
		for (i in 1:a) for (j in 1:b) {
			L0_star[i, j] <- n_ij * (
			-lgamma(delta_0[i, j]) -
			delta_0[i, j] * log(beta_0) -
			x_bar[i, j] / beta_0 +
			(delta_0[i, j] - 1) * log(x_tilde[i, j])
			)
		}
		L2 <- sum(L0_star)
		
		# 7. Likelihood-ratio statistic
		lambda <- exp(L2 - L1)
		lambda_star <- -2 * log(lambda)
		
		return(lambda_star)
	}
	
	# ============================================================
	# Monte Carlo simulation
	# ============================================================
	
	start.time <- Sys.time()
	
	for (q in 1:Q) {
		
		count <- 0
		
		# Generate observed data
		repeat {
			for (i in 1:a) for (j in 1:b) for (k in 1:n_ij) {
				x[i, j, k] <- rgamma(
				1, shape = delta_in[i, j], scale = beta
				)
			}
			
			lambda_star <- cal_lambda(a, b, n_ij, x)
			
			if (!is.na(lambda_star) && !is.infinite(lambda_star))
			break
		}
		
		# Parametric bootstrap
		lambda_rep <- rep(NA_real_, M)
		
		for (m in 1:M) {
			repeat {
				xstar <- array(NA_real_, c(a, b, n_ij))
				
				for (i in 1:a) for (j in 1:b) for (k in 1:n_ij) {
					xstar[i, j, k] <- rgamma(
					1, shape = delta_gen[i, j], scale = beta_gen
					)
				}
				
				resultm <- cal_lambda(a, b, n_ij, xstar)
				
				if (!is.na(resultm) && !is.infinite(resultm))
				break
			}
			
			lambda_rep[m] <- resultm
		}
		
		# Bootstrap upper critical value
		lambda_upper <- as.numeric(
		quantile(lambda_rep, probs = 1 - alpha,
		type = 7, na.rm = TRUE)
		)
		
		# ALRT
		if (lambda_star > qchisq(1 - alpha, df = a * b - 1))
		count3 <- count3 + 1
		
		# PBLRT
		if (lambda_star > lambda_upper)
		count2 <- count2 + 1
	}
	
	# ============================================================
	# Results
	# ============================================================
	
	cat("Q =", Q, "M =", M, "a =", a, "b =", b,
	"alpha =", alpha, "n_ij =", n_ij,
	"delta1 =", delta1, "delta2 =", delta2,
	"delta3 =", delta3, "delta4 =", delta4, "\n")
	
	cat("ALRT count3 =", count3, "\n")
	cat("PBLRT count2 =", count2, "\n")
	
	p_ALRT <- count3 / Q
	p_PBLRT <- count2 / Q
	
	cat("p_ALRT =", p_ALRT, "\n")
	cat("p_PBLRT =", p_PBLRT, "\n")
	
	# Computing time
	end.time <- Sys.time()
	time.taken <- end.time - start.time
	print(time.taken)
	
	# ============================================================
	# End of script
	# ============================================================
	
\end{lstlisting}
\section{The codes of the simulation Problem - 1}
\begin{lstlisting}[style=Rcompact,
	caption={Simulation code for Table 3.1},
	label={lst:table31}]
	
	# ============================================================
	# Simulation for Table 3.1
	# ============================================================
	
	rm(list = ls(all = TRUE))
	library(nleqslv)
	
	myseed <- 123456
	set.seed(myseed)
	
	# Counters
	count  <- 0  # First call within each q
	count2 <- 0  # PBLRT rejections
	count3 <- 0  # ALRT rejections
	
	# Simulation settings
	a <- b <- 2
	n_ij <- 5
	alpha <- 0.01
	delta1 <- 1
	delta2 <- 0.5
	Q <- M <- 5000
	beta_in <- 1
	
	# Data and bootstrap parameter containers
	x <- array(NA_real_, c(a, b, n_ij))
	delta_in <- matrix(c(delta1, delta2, delta1, delta2), a, b)
	delta_igen <- numeric(a)
	beta_gen <- numeric(1)
	
	# ============================================================
	# Function to calculate lambda_star
	# ============================================================
	
	cal_lambda <- function(a, b, n_ij, x) {
		
		# 1. Arithmetic and geometric means
		x_bar <- matrix(NA_real_, a, b)
		x_tilde <- matrix(NA_real_, a, b)
		
		for (i in 1:a) for (j in 1:b) {
			xx <- x[i, j, ]
			x_bar[i, j] <- mean(xx, na.rm = TRUE)
			x_tilde[i, j] <- exp(mean(log(xx), na.rm = TRUE))
		}
		
		# 2. Unrestricted MLEs: Equation (3.6)
		solve_system1 <- function(n_ij, x_bar, x_tilde) {
			
			i_size <- nrow(x_bar)
			j_size <- ncol(x_bar)
			
			equations1 <- function(delta_tam) {
				deltaM <- matrix(delta_tam, nrow = i_size,
				ncol = j_size)
				eqs <- matrix(NA_real_, nrow = i_size,
				ncol = j_size)
				
				s1 <- sum(n_ij * deltaM)
				s2 <- sum(n_ij * x_bar)
				
				for (ii in 1:i_size) for (jj in 1:j_size) {
					eqs[ii, jj] <- log(s1) -
					digamma(deltaM[ii, jj]) -
					log(s2 / x_tilde[ii, jj])
				}
				
				as.vector(eqs)
			}
			
			initial_num <- runif(1, 0.01, 1)
			initial_guess <- rep(initial_num, i_size * j_size)
			
			sol <- try(
			nleqslv::nleqslv(
			initial_guess, equations1,
			method = "Newton",
			control = list(maxit = 2000, trace = 0)
			),
			silent = TRUE
			)
			
			if (inherits(sol, "try-error")) return(NA)
			if (is.null(sol$termcd) || sol$termcd != 1) return(NA)
			if (any(sol$x <= 0)) return(NA)
			
			matrix(sol$x, nrow = i_size, ncol = j_size)
		}
		
		delta <- solve_system1(n_ij, x_bar, x_tilde)
		if (any(is.na(delta))) return(NA_real_)
		
		beta <- sum(n_ij * x_bar) / sum(n_ij * delta)
		
		# 3. Log-likelihood L1: Equation (3.4)
		L_star <- matrix(0, a, b)
		
		for (i in 1:a) for (j in 1:b) {
			L_star[i, j] <- n_ij * (
			-lgamma(delta[i, j]) -
			delta[i, j] * log(beta) -
			x_bar[i, j] / beta +
			(delta[i, j] - 1) * log(x_tilde[i, j])
			)
		}
		
		L1 <- sum(L_star)
		
		# 4. Restricted MLEs: Equation (3.12)
		solve_system2 <- function(n_ij, x_bar, x_tilde) {
			
			i_size <- nrow(x_bar)
			b_size <- ncol(x_bar)
			
			equations2 <- function(deltam2) {
				eqs2 <- numeric(i_size)
				
				# Balanced case: n_i = sum_j n_ij and v_ij = n_ij / n_i
				ni <- b_size * n_ij
				vij <- n_ij / ni
				
				t1 <- sum(ni * deltam2)
				t2 <- sum(n_ij * x_bar)
				
				for (ii in 1:i_size) {
					eqs2[ii] <- log(t1) -
					b_size * vij * digamma(deltam2[ii]) -
					log(t2 / prod(x_tilde[ii, ]^vij))
				}
				
				eqs2
			}
			
			initial_num2 <- runif(1, 0.01, 1)
			initial_guess2 <- rep(initial_num2, i_size)
			
			sol2 <- try(
			nleqslv::nleqslv(
			initial_guess2, equations2,
			method = "Newton",
			control = list(maxit = 2000, trace = 0)
			),
			silent = TRUE
			)
			
			if (inherits(sol2, "try-error")) return(NA)
			if (is.null(sol2$termcd) || sol2$termcd != 1) return(NA)
			if (any(sol2$x <= 0)) return(NA)
			
			sol2$x
		}
		
		delta_i0 <- solve_system2(n_ij, x_bar, x_tilde)
		if (any(is.na(delta_i0))) return(NA_real_)
		
		# 5. Calculate beta_0: Equation (3.13)
		beta_0 <- sum(n_ij * x_bar) /
		sum(b * n_ij * delta_i0)
		
		# Store estimates from the first call within each q
		if (count == 0) {
			delta_igen <<- delta_i0
			beta_gen <<- beta_0
			count <<- count + 1
		}
		
		# 6. Log-likelihood L2: Equation (3.9)
		L0_star <- matrix(NA_real_, a, b)
		
		for (i in 1:a) for (j in 1:b) {
			L0_star[i, j] <- n_ij * (
			-lgamma(delta_i0[i]) -
			delta_i0[i] * log(beta_0) -
			x_bar[i, j] / beta_0 +
			(delta_i0[i] - 1) * log(x_tilde[i, j])
			)
		}
		
		L2 <- sum(L0_star)
		
		# 7. Likelihood-ratio statistic
		lambda <- exp(L2 - L1)
		lambda_star <- -2 * log(lambda)
		
		return(lambda_star)
	}
	
	# ============================================================
	# Monte Carlo simulation
	# ============================================================
	
	start.time <- Sys.time()
	
	for (q in 1:Q) {
		
		count <- 0
		
		# Step 1: Generate observed data
		repeat {
			for (i in 1:a) for (j in 1:b) for (k in 1:n_ij) {
				x[i, j, k] <- rgamma(
				1, shape = delta_in[i, j], scale = beta_in
				)
			}
			
			lambda_star <- cal_lambda(a, b, n_ij, x)
			
			if (!is.na(lambda_star) && !is.infinite(lambda_star))
			break
		}
		
		# Step 2: Parametric bootstrap
		lambda_rep <- rep(NA_real_, M)
		
		for (m in 1:M) {
			repeat {
				xstar <- array(NA_real_, c(a, b, n_ij))
				
				for (i in 1:a) for (j in 1:b) for (k in 1:n_ij) {
					xstar[i, j, k] <- rgamma(
					1, shape = delta_igen[i], scale = beta_gen
					)
				}
				
				resultm <- cal_lambda(a, b, n_ij, xstar)
				
				if (!is.na(resultm) && !is.infinite(resultm))
				break
			}
			
			lambda_rep[m] <- resultm
		}
		
		# Step 3: Bootstrap critical value
		lambda_upper <- as.numeric(
		quantile(lambda_rep, probs = 1 - alpha,
		type = 7, na.rm = TRUE)
		)
		
		# ALRT
		if (lambda_star > qchisq(1 - alpha, df = a * (b - 1)))
		count3 <- count3 + 1
		
		# PBLRT
		if (lambda_star > lambda_upper)
		count2 <- count2 + 1
	}
	
	# ============================================================
	# Results
	# ============================================================
	
	cat("Q =", Q, "M =", M, "a =", a, "b =", b,
	"alpha =", alpha, "n_ij =", n_ij,
	"delta1 =", delta1, "delta2 =", delta2, "\n")
	
	cat("ALRT count3 =", count3, "\n")
	cat("PBLRT count2 =", count2, "\n")
	
	p_ALRT <- count3 / Q
	p_PBLRT <- count2 / Q
	
	cat("p_ALRT =", p_ALRT, "\n")
	cat("p_PBLRT =", p_PBLRT, "\n")
	
	end.time <- Sys.time()
	time.taken <- end.time - start.time
	print(time.taken)
	
	# ============================================================
	# End of script
	# ============================================================
	
\end{lstlisting}
\begin{lstlisting}[style=Rcompact,
	caption={Simulation code for Table 3.2},
	label={lst:table32}]
	
	# ============================================================
	# Simulation for Table 3.2
	# ============================================================
	
	rm(list = ls(all = TRUE))
	library(nleqslv)
	
	myseed <- 123456
	set.seed(myseed)
	
	# Counters
	count  <- 0  # First call within each q
	count2 <- 0  # PBLRT rejections
	count3 <- 0  # ALRT rejections
	
	# Simulation settings
	a <- b <- 2
	n_ij <- 5
	alpha <- 0.01
	delta1 <- 1
	delta2 <- 0.5
	Q <- M <- 5000
	
	# Data and bootstrap parameter containers
	x <- array(NA_real_, c(a, b, n_ij))
	delta_in <- matrix(c(delta1, delta2, delta1, delta2), a, b)
	delta_gen <- matrix(NA_real_, a, b)
	mu_gen <- numeric(a)
	
	# ============================================================
	# Root-bracketing function
	# ============================================================
	
	bracket_root <- function(f, lower = .Machine$double.eps,
	upper = 5e4, max_expand = 8) {
		
		f_low <- suppressWarnings(
		tryCatch(f(lower), error = function(e) NA_real_)
		)
		f_up <- suppressWarnings(
		tryCatch(f(upper), error = function(e) NA_real_)
		)
		
		if (!is.na(f_low) && !is.na(f_up) && f_low * f_up <= 0)
		return(c(lower, upper))
		
		# Expand the upper bound
		u <- upper
		for (k in 1:max_expand) {
			u <- u * 10
			f_up <- suppressWarnings(
			tryCatch(f(u), error = function(e) NA_real_)
			)
			
			if (!is.na(f_low) && !is.na(f_up) && f_low * f_up <= 0)
			return(c(lower, u))
		}
		
		# Reduce the lower bound
		l <- lower
		for (k in 1:max_expand) {
			l <- l / 10
			if (l <= .Machine$double.eps)
			l <- .Machine$double.eps
			
			f_low <- suppressWarnings(
			tryCatch(f(l), error = function(e) NA_real_)
			)
			
			if (!is.na(f_low) && !is.na(f_up) && f_low * f_up <= 0)
			return(c(l, u))
		}
		
		return(NA)
	}
	
	# ============================================================
	# Function to calculate lambda_star
	# ============================================================
	
	cal_lambda <- function(a, b, n_ij, x) {
		
		# Arithmetic and geometric means
		x_bar <- matrix(NA_real_, a, b)
		x_tilde <- matrix(NA_real_, a, b)
		
		for (i in 1:a) for (j in 1:b) {
			xx <- x[i, j, ]
			x_bar[i, j] <- mean(xx, na.rm = TRUE)
			x_tilde[i, j] <- exp(mean(log(xx), na.rm = TRUE))
		}
		
		# 1. Unrestricted MLEs: Equation (3.18)
		delta <- matrix(NA_real_, a, b)
		beta <- matrix(NA_real_, a, b)
		
		for (i in 1:a) for (j in 1:b) {
			f1 <- function(a_) {
				digamma(a_) - log(a_) -
				log(x_tilde[i, j] / x_bar[i, j])
			}
			
			br <- bracket_root(
			f1, lower = .Machine$double.eps, upper = 5e4
			)
			if (all(is.na(br))) return(NA_real_)
			
			root <- suppressWarnings(
			tryCatch(
			uniroot(f1, interval = br)$root,
			error = function(e) NA_real_
			)
			)
			if (is.na(root)) return(NA_real_)
			
			delta[i, j] <- root
			beta[i, j] <- x_bar[i, j] / delta[i, j]
		}
		
		# 2. Log-likelihood L1: Equation (3.16)
		L_star <- matrix(0, a, b)
		
		for (i in 1:a) for (j in 1:b) {
			L_star[i, j] <- n_ij * (
			-lgamma(delta[i, j]) -
			delta[i, j] * log(beta[i, j]) -
			x_bar[i, j] / beta[i, j] +
			(delta[i, j] - 1) * log(x_tilde[i, j])
			)
		}
		
		L1 <- sum(L_star)
		
		# 3. Restricted MLEs: Equation (3.21)
		solve_system <- function(n_ij, x_bar, x_tilde) {
			
			i_size <- nrow(x_bar)
			j_size <- ncol(x_bar)
			
			equations <- function(delta_tam) {
				deltaM <- matrix(delta_tam, nrow = i_size,
				ncol = j_size)
				eqs <- matrix(NA_real_, nrow = i_size,
				ncol = j_size)
				
				for (ii in 1:i_size) {
					
					# Quantities used to estimate mu_i
					num <- sum(
					n_ij * deltaM[ii, ] * x_bar[ii, ]
					)
					denom <- sum(n_ij * deltaM[ii, ])
					
					# Numerical protection
					num <- max(num, .Machine$double.xmin)
					denom <- max(denom, .Machine$double.xmin)
					
					for (jj in 1:j_size) {
						eqs[ii, jj] <- digamma(deltaM[ii, jj]) +
						log(num / denom) +
						x_bar[ii, jj] * (denom / num) - 1 -
						log(max(
						deltaM[ii, jj] * x_tilde[ii, jj],
						.Machine$double.xmin
						))
					}
				}
				
				as.vector(eqs)
			}
			
			# Initial values
			initial_num <- runif(1, 0.001, 1)
			initial_guess <- rep(initial_num, i_size * j_size)
			
			# Solve using Broyden's method
			sol <- try(
			nleqslv::nleqslv(
			initial_guess, equations,
			method = "Broyden",
			control = list(maxit = 2000, trace = 0)
			),
			silent = TRUE
			)
			
			if (inherits(sol, "try-error")) return(NA)
			if (is.null(sol$termcd) || sol$termcd != 1) return(NA)
			
			matrix(sol$x, nrow = i_size, ncol = j_size)
		}
		
		delta_0 <- solve_system(n_ij, x_bar, x_tilde)
		if (any(is.na(delta_0))) return(NA_real_)
		
		# 4. Estimate mu_i: Equation (3.20)
		mu <- numeric(a)
		
		for (i in 1:a) {
			mu[i] <- sum(
			n_ij * delta_0[i, ] * x_bar[i, ]
			) / sum(n_ij * delta_0[i, ])
		}
		
		# Store estimates from the first call within each q
		if (count == 0) {
			delta_gen <<- delta_0
			mu_gen <<- mu
			count <<- count + 1
		}
		
		# 5. Log-likelihood L2: Equation (3.19)
		L0_star <- matrix(NA_real_, a, b)
		
		for (i in 1:a) for (j in 1:b) {
			beta_0_ij <- mu[i] / delta_0[i, j]
			
			L0_star[i, j] <- n_ij * (
			-lgamma(delta_0[i, j]) -
			delta_0[i, j] * log(beta_0_ij) -
			x_bar[i, j] / beta_0_ij +
			(delta_0[i, j] - 1) * log(x_tilde[i, j])
			)
		}
		
		L2 <- sum(L0_star)
		
		# 6. Likelihood-ratio statistic
		lambda <- exp(L2 - L1)
		lambda_star <- -2 * log(lambda)
		
		return(lambda_star)
	}
	
	# ============================================================
	# Monte Carlo simulation
	# ============================================================
	
	start.time <- Sys.time()
	
	for (q in 1:Q) {
		
		# Reset first-call indicator
		count <- 0
		
		# Step 1: Generate observed data
		repeat {
			for (i in 1:a) for (j in 1:b) for (k in 1:n_ij) {
				x[i, j, k] <- rgamma(
				1, shape = delta_in[i, j], scale = 1
				)
			}
			
			lambda_star <- cal_lambda(a, b, n_ij, x)
			
			if (!is.na(lambda_star) && !is.infinite(lambda_star))
			break
		}
		
		# Step 2: Parametric bootstrap
		lambda_rep <- rep(NA_real_, M)
		
		for (m in 1:M) {
			repeat {
				xstar <- array(NA_real_, c(a, b, n_ij))
				
				# Generate bootstrap samples
				for (i in 1:a) for (j in 1:b) {
					beta_ij <- mu_gen[i] / delta_gen[i, j]
					
					for (k in 1:n_ij) {
						xstar[i, j, k] <- rgamma(
						1, shape = delta_gen[i, j],
						scale = beta_ij
						)
					}
				}
				
				resultm <- cal_lambda(a, b, n_ij, xstar)
				
				if (!is.na(resultm) && !is.infinite(resultm))
				break
			}
			
			lambda_rep[m] <- resultm
		}
		
		# Step 3: Bootstrap critical value
		lambda_upper <- as.numeric(
		quantile(lambda_rep, probs = 1 - alpha,
		type = 7, na.rm = TRUE)
		)
		
		# Step 4: ALRT and PBLRT
		# Degrees of freedom:
		# unrestricted model: 2ab parameters
		# restricted model: ab + a parameters
		df_LRT <- a * (b - 1)
		
		# Asymptotic likelihood-ratio test
		if (lambda_star > qchisq(1 - alpha, df = df_LRT))
		count3 <- count3 + 1
		
		# Parametric bootstrap likelihood-ratio test
		if (lambda_star > lambda_upper)
		count2 <- count2 + 1
	}
	
	# ============================================================
	# Results
	# ============================================================
	
	cat("Q =", Q, "M =", M, "a =", a, "b =", b,
	"alpha =", alpha, "n_ij =", n_ij,
	"delta1 =", delta1, "delta2 =", delta2, "\n")
	
	cat("ALRT count3 =", count3, "\n")
	cat("PBLRT count2 =", count2, "\n")
	
	p_ALRT <- count3 / Q
	p_PBLRT <- count2 / Q
	
	cat("p_ALRT =", p_ALRT, "\n")
	cat("p_PBLRT =", p_PBLRT, "\n")
	
	end.time <- Sys.time()
	time.taken <- end.time - start.time
	print(time.taken)
	
	# ============================================================
	# End of script
	# ============================================================
	
\end{lstlisting}
\section{The codes of the simulation Problem - 3}
\begin{lstlisting}[style=Rcompact,
	caption={Simulation code for Table 4.1},
	label={lst:table41}]
	# ============================================================
	# Simulation for Table 4.1
	# ============================================================
	rm(list = ls(all = TRUE))
	library(nleqslv)
	set.seed(123456)
	# Simulation settings
	a <- b <- 2
	n_ij <- 5
	alpha <- 0.01
	delta_in <- 1
	beta_in <- 1
	Q <- M <- 5000
	# Counters
	count1 <- 0                         # First call within each q
	count2 <- 0                         # PBLRT rejections
	count3 <- 0                         # ALRT rejections
	# Data and bootstrap parameter containers
	x <- array(NA_real_, c(a, b, n_ij))
	delta_gen <- NA_real_
	beta_gen <- NA_real_
	# ============================================================
	# Root-bracketing function
	# ============================================================
	bracket_root <- function(
	f, lower = .Machine$double.eps,
	upper = 5e4, max_expand = 8
	) {
		f_low <- suppressWarnings(
		tryCatch(f(lower), error = function(e) NA_real_)
		)
		f_up <- suppressWarnings(
		tryCatch(f(upper), error = function(e) NA_real_)
		)
		if (!is.na(f_low) && !is.na(f_up) && f_low * f_up <= 0)
		return(c(lower, upper))
		# Expand the upper bound
		u <- upper
		for (k in 1:max_expand) {
			u <- u * 10
			f_up <- suppressWarnings(
			tryCatch(f(u), error = function(e) NA_real_)
			)
			if (!is.na(f_low) && !is.na(f_up) && f_low * f_up <= 0)
			return(c(lower, u))
		}
		# Reduce the lower bound
		l <- lower
		for (k in 1:max_expand) {
			l <- max(l / 10, .Machine$double.eps)
			f_low <- suppressWarnings(
			tryCatch(f(l), error = function(e) NA_real_)
			)
			if (!is.na(f_low) && !is.na(f_up) && f_low * f_up <= 0)
			return(c(l, u))
		}
		return(NA)
	}
	# ============================================================
	# Function to calculate the likelihood-ratio statistic
	# ============================================================
	cal_lambda <- function(a, b, n_ij, x) {
		# Arithmetic and geometric means
		x_bar <- matrix(NA_real_, a, b)
		x_tilde <- matrix(NA_real_, a, b)
		for (i in 1:a) for (j in 1:b) {
			xx <- x[i, j, ]
			x_bar[i, j] <- mean(xx)
			x_tilde[i, j] <- exp(mean(log(xx)))
		}
		# ==========================================================
		# 1. Unrestricted MLEs: Equations (3.6) and (3.7)
		# ==========================================================
		solve_system1 <- function(x_bar, x_tilde) {
			i_size <- nrow(x_bar)
			j_size <- ncol(x_bar)
			equations1 <- function(delta_tam) {
				deltaM <- matrix(
				delta_tam,
				nrow = i_size,
				ncol = j_size
				)
				eqs <- matrix(NA_real_, i_size, j_size)
				s1 <- sum(n_ij * deltaM)
				s2 <- sum(n_ij * x_bar)
				for (ii in 1:i_size) for (jj in 1:j_size) {
					eqs[ii, jj] <-
					log(s1) -
					digamma(deltaM[ii, jj]) -
					log(s2 / x_tilde[ii, jj])
				}
				as.vector(eqs)
			}
			initial_guess <- rep(
			runif(1, 0.001, 1),
			i_size * j_size
			)
			sol <- try(
			nleqslv::nleqslv(
			initial_guess,
			equations1,
			method = "Newton",
			control = list(maxit = 2000, trace = 0)
			),
			silent = TRUE
			)
			if (inherits(sol, "try-error") ||
			is.null(sol$termcd) ||
			sol$termcd != 1 ||
			any(sol$x <= 0))
			return(NA)
			matrix(sol$x, i_size, j_size)
		}
		delta <- solve_system1(x_bar, x_tilde)
		if (any(is.na(delta)))
		return(NA_real_)
		beta <-
		sum(n_ij * x_bar) /
		sum(n_ij * delta)
		# ==========================================================
		# 2. Unrestricted log-likelihood: Equation (3.4)
		# ==========================================================
		L1 <- 0
		for (i in 1:a) for (j in 1:b) {
			L1 <- L1 + n_ij * (
			-lgamma(delta[i, j]) -
			delta[i, j] * log(beta) -
			x_bar[i, j] / beta +
			(delta[i, j] - 1) *
			log(x_tilde[i, j])
			)
		}
		# ==========================================================
		# 3. Restricted MLEs: Equations (4.6) and (4.7)
		# ==========================================================
		f1 <- function(delta_0) {
			digamma(delta_0) -
			log(delta_0 * a * b * n_ij) -
			sum(n_ij * log(x_tilde)) /
			(a * b * n_ij) +
			log(sum(n_ij * x_bar))
		}
		br <- bracket_root(f1)
		if (all(is.na(br)))
		return(NA_real_)
		delta_0 <- suppressWarnings(
		tryCatch(
		uniroot(f1, interval = br)$root,
		error = function(e) NA_real_
		)
		)
		if (is.na(delta_0))
		return(NA_real_)
		beta_0 <-
		sum(n_ij * x_bar) /
		(delta_0 * a * b * n_ij)
		# Store the restricted estimates from the observed sample
		if (count1 == 0) {
			delta_gen <<- delta_0
			beta_gen <<- beta_0
			count1 <<- count1 + 1
		}
		# ==========================================================
		# 4. Restricted log-likelihood: Equation (4.3)
		# ==========================================================
		L2 <- 0
		for (i in 1:a) for (j in 1:b) {
			L2 <- L2 + n_ij * (
			-lgamma(delta_0) -
			delta_0 * log(beta_0) -
			x_bar[i, j] / beta_0 +
			(delta_0 - 1) *
			log(x_tilde[i, j])
			)
		}
		# ==========================================================
		# 5. Likelihood-ratio statistic
		# ==========================================================
		lambda_star <- 2 * (L1 - L2)
		return(lambda_star)
	}
	# ============================================================
	# Monte Carlo simulation
	# ============================================================
	start.time <- Sys.time()
	for (q in 1:Q) {
		# Reset the first-call indicator
		count1 <- 0
		if (q %% 100 == 0)
		cat("q =", q, "\n")
		# ==========================================================
		# Step 1: Generate the observed sample
		# ==========================================================
		repeat {
			for (i in 1:a) for (j in 1:b) {
				x[i, j, ] <- rgamma(
				n_ij,
				shape = delta_in,
				scale = beta_in
				)
			}
			lambda_star <- cal_lambda(a, b, n_ij, x)
			if (!is.na(lambda_star) &&
			!is.infinite(lambda_star))
			break
		}
		# ==========================================================
		# Step 2: Parametric bootstrap
		# ==========================================================
		lambda_rep <- numeric(M)
		for (m in 1:M) {
			repeat {
				xstar <- array(
				NA_real_,
				c(a, b, n_ij)
				)
				for (i in 1:a) for (j in 1:b) {
					xstar[i, j, ] <- rgamma(
					n_ij,
					shape = delta_gen,
					scale = beta_gen
					)
				}
				resultm <- cal_lambda(
				a, b, n_ij, xstar
				)
				if (!is.na(resultm) &&
				!is.infinite(resultm))
				break
			}
			lambda_rep[m] <- resultm
		}
		# ==========================================================
		# Step 3: Bootstrap critical value
		# ==========================================================
		lambda_upper <- as.numeric(
		quantile(
		lambda_rep,
		probs = 1 - alpha,
		type = 7,
		na.rm = TRUE
		)
		)
		# ==========================================================
		# Step 4: ALRT and PBLRT
		# ==========================================================
		# Number of restrictions:
		# unrestricted: ab shape parameters + 1 common scale
		# restricted: 1 shape parameter + 1 common scale
		df_LRT <- a * b - 1
		# Asymptotic likelihood-ratio test
		if (lambda_star >
		qchisq(1 - alpha, df = df_LRT))
		count3 <- count3 + 1
		# Parametric bootstrap likelihood-ratio test
		if (lambda_star > lambda_upper)
		count2 <- count2 + 1
	}
	# ============================================================
	# Results
	# ============================================================
	p_ALRT <- count3 / Q
	p_PBLRT <- count2 / Q
	cat(
	"Q =", Q,
	"M =", M,
	"a =", a,
	"b =", b,
	"alpha =", alpha,
	"n_ij =", n_ij,
	"delta =", delta_in,
	"\n"
	)
	cat("ALRT count =", count3, "\n")
	cat("PBLRT count =", count2, "\n")
	cat("p_ALRT =", p_ALRT, "\n")
	cat("p_PBLRT =", p_PBLRT, "\n")
	end.time <- Sys.time()
	time.taken <- end.time - start.time
	print(time.taken)
	# ============================================================
	# End of script
	# ============================================================
\end{lstlisting}
\begin{lstlisting}[style=Rcompact,
	caption={Simulation code for Table 4.2},
	label={lst:table42}]
	# ============================================================
	# Simulation for Table 4.2
	# ============================================================
	rm(list = ls(all = TRUE))
	library(nleqslv)
	set.seed(123456)
	# ============================================================
	# Simulation settings
	# ============================================================
	a <- b <- 2
	n_ij <- 5
	alpha <- 0.01
	delta_in <- 1
	beta_in <- 1
	Q <- M <- 5000
	# ============================================================
	# Counters
	# ============================================================
	count1 <- 0                    # First call within each q
	count2 <- 0                    # PBLRT rejections
	count3 <- 0                    # ALRT rejections
	# ============================================================
	# Data and bootstrap parameter containers
	# ============================================================
	x <- array(NA_real_, c(a, b, n_ij))
	delta_gen <- matrix(NA_real_, a, b)
	beta_gen <- matrix(NA_real_, a, b)
	# ============================================================
	# Root-bracketing function
	# ============================================================
	bracket_root <- function(
	f,
	lower = .Machine$double.eps,
	upper = 5e4,
	max_expand = 8
	) {
		f_low <- suppressWarnings(
		tryCatch(f(lower), error = function(e) NA_real_)
		)
		f_up <- suppressWarnings(
		tryCatch(f(upper), error = function(e) NA_real_)
		)
		if (!is.na(f_low) &&
		!is.na(f_up) &&
		f_low * f_up <= 0)
		return(c(lower, upper))
		# Expand the upper bound
		u <- upper
		for (k in 1:max_expand) {
			u <- u * 10
			f_up <- suppressWarnings(
			tryCatch(f(u), error = function(e) NA_real_)
			)
			if (!is.na(f_low) &&
			!is.na(f_up) &&
			f_low * f_up <= 0)
			return(c(lower, u))
		}
		# Reduce the lower bound
		l <- lower
		for (k in 1:max_expand) {
			l <- max(
			l / 10,
			.Machine$double.eps
			)
			f_low <- suppressWarnings(
			tryCatch(f(l), error = function(e) NA_real_)
			)
			if (!is.na(f_low) &&
			!is.na(f_up) &&
			f_low * f_up <= 0)
			return(c(l, u))
		}
		return(NA)
	}
	# ============================================================
	# Function to calculate the likelihood-ratio statistic
	# ============================================================
	cal_lambda <- function(a, b, n_ij, x) {
		# ----------------------------------------------------------
		# Arithmetic and geometric means
		# ----------------------------------------------------------
		x_bar <- matrix(NA_real_, a, b)
		x_tilde <- matrix(NA_real_, a, b)
		for (i in 1:a) for (j in 1:b) {
			xx <- x[i, j, ]
			x_bar[i, j] <- mean(xx)
			x_tilde[i, j] <-
			exp(mean(log(xx)))
		}
		# ==========================================================
		# 1. Unrestricted MLEs: Equation (3.18)
		# ==========================================================
		delta <- matrix(NA_real_, a, b)
		beta <- matrix(NA_real_, a, b)
		for (i in 1:a) for (j in 1:b) {
			f1 <- function(delta_ij) {
				digamma(delta_ij) -
				log(delta_ij) -
				log(x_tilde[i, j] / x_bar[i, j])
			}
			br <- bracket_root(f1)
			if (all(is.na(br)))
			return(NA_real_)
			root <- suppressWarnings(
			tryCatch(
			uniroot(f1, interval = br)$root,
			error = function(e) NA_real_
			)
			)
			if (is.na(root))
			return(NA_real_)
			delta[i, j] <- root
			beta[i, j] <-
			x_bar[i, j] / delta[i, j]
		}
		# ==========================================================
		# 2. Unrestricted log-likelihood: Equation (3.16)
		# ==========================================================
		L1 <- 0
		for (i in 1:a) for (j in 1:b) {
			L1 <- L1 + n_ij * (
			-lgamma(delta[i, j]) -
			delta[i, j] *
			log(beta[i, j]) -
			x_bar[i, j] /
			beta[i, j] +
			(delta[i, j] - 1) *
			log(x_tilde[i, j])
			)
		}
		# ==========================================================
		# 3. Restricted MLEs: Equations (4.12) and (4.13)
		# ==========================================================
		solve_system <- function(x_bar, x_tilde) {
			i_size <- nrow(x_bar)
			j_size <- ncol(x_bar)
			equations <- function(delta_tam) {
				deltaM <- matrix(
				delta_tam,
				nrow = i_size,
				ncol = j_size
				)
				eqs <- matrix(
				NA_real_,
				i_size,
				j_size
				)
				for (ii in 1:i_size) {
					for (jj in 1:j_size) {
						# Quantities used to estimate the common mean
						num <-
						n_ij *
						deltaM[ii, jj] *
						x_bar[ii, jj]
						denom <-
						n_ij *
						deltaM[ii, jj]
						# Numerical protection
						num <- max(
						num,
						.Machine$double.xmin
						)
						denom <- max(
						denom,
						.Machine$double.xmin
						)
						eqs[ii, jj] <-
						digamma(deltaM[ii, jj]) +
						log(num / denom) +
						x_bar[ii, jj] *
						(denom / num) -
						1 -
						log(
						max(
						deltaM[ii, jj] *
						x_tilde[ii, jj],
						.Machine$double.xmin
						)
						)
					}
				}
				as.vector(eqs)
			}
			initial_guess <- rep(
			runif(1, 0.001, 1),
			i_size * j_size
			)
			sol <- try(
			nleqslv::nleqslv(
			initial_guess,
			equations,
			method = "Broyden",
			control = list(
			maxit = 2000,
			trace = 0
			)
			),
			silent = TRUE
			)
			if (inherits(sol, "try-error") ||
			is.null(sol$termcd) ||
			sol$termcd != 1)
			return(NA)
			matrix(
			sol$x,
			nrow = i_size,
			ncol = j_size
			)
		}
		delta_0 <- solve_system(
		x_bar,
		x_tilde
		)
		if (any(is.na(delta_0)))
		return(NA_real_)
		# Common mean: Equation (4.13)
		mu <-
		sum(
		n_ij *
		delta_0 *
		x_bar
		) /
		sum(
		n_ij *
		delta_0
		)
		# Restricted scale parameters
		beta_0 <- mu / delta_0
		# Store estimates from the observed sample
		if (count1 == 0) {
			delta_gen <<- delta_0
			beta_gen <<- beta_0
			count1 <<- count1 + 1
		}
		# ==========================================================
		# 4. Restricted log-likelihood: Equation (4.11)
		# ==========================================================
		L2 <- 0
		for (i in 1:a) for (j in 1:b) {
			L2 <- L2 + n_ij * (
			-lgamma(delta_0[i, j]) -
			delta_0[i, j] *
			log(beta_0[i, j]) -
			x_bar[i, j] /
			beta_0[i, j] +
			(delta_0[i, j] - 1) *
			log(x_tilde[i, j])
			)
		}
		# ==========================================================
		# 5. Likelihood-ratio statistic
		# ==========================================================
		lambda_star <- 2 * (L1 - L2)
		return(lambda_star)
	}
	# ============================================================
	# Monte Carlo simulation
	# ============================================================
	start.time <- Sys.time()
	for (q in 1:Q) {
		# Reset the first-call indicator
		count1 <- 0
		# ==========================================================
		# Step 1: Generate the observed sample
		# ==========================================================
		repeat {
			for (i in 1:a) for (j in 1:b) {
				x[i, j, ] <- rgamma(
				n_ij,
				shape = delta_in,
				scale = beta_in
				)
			}
			lambda_star <- cal_lambda(
			a, b, n_ij, x
			)
			if (!is.na(lambda_star) &&
			!is.infinite(lambda_star))
			break
		}
		# ==========================================================
		# Step 2: Parametric bootstrap
		# ==========================================================
		lambda_rep <- numeric(M)
		for (m in 1:M) {
			repeat {
				xstar <- array(
				NA_real_,
				c(a, b, n_ij)
				)
				for (i in 1:a) for (j in 1:b) {
					xstar[i, j, ] <- rgamma(
					n_ij,
					shape = delta_gen[i, j],
					scale = beta_gen[i, j]
					)
				}
				resultm <- cal_lambda(
				a, b, n_ij, xstar
				)
				if (!is.na(resultm) &&
				!is.infinite(resultm))
				break
			}
			lambda_rep[m] <- resultm
		}
		# ==========================================================
		# Step 3: Bootstrap critical value
		# ==========================================================
		lambda_upper <- as.numeric(
		quantile(
		lambda_rep,
		probs = 1 - alpha,
		type = 7,
		na.rm = TRUE
		)
		)
		# ==========================================================
		# Step 4: ALRT and PBLRT
		# ==========================================================
		# Number of restrictions
		df_LRT <- a * b - 1
		# Asymptotic likelihood-ratio test
		if (lambda_star >
		qchisq(1 - alpha, df = df_LRT))
		count3 <- count3 + 1
		# Parametric bootstrap likelihood-ratio test
		if (lambda_star > lambda_upper)
		count2 <- count2 + 1
	}
	# ============================================================
	# Results
	# ============================================================
	p_ALRT <- count3 / Q
	p_PBLRT <- count2 / Q
	cat(
	"Q =", Q,
	"M =", M,
	"a =", a,
	"b =", b,
	"alpha =", alpha,
	"n_ij =", n_ij,
	"delta =", delta_in,
	"\n"
	)
	cat("ALRT count =", count3, "\n")
	cat("PBLRT count =", count2, "\n")
	cat("p_ALRT =", p_ALRT, "\n")
	cat("p_PBLRT =", p_PBLRT, "\n")
	end.time <- Sys.time()
	time.taken <- end.time - start.time
	print(time.taken)
	# ============================================================
	# End of script
	# ============================================================
\end{lstlisting}
\section{The codes of the simulation Problem - 4}
\begin{lstlisting}[style=Rcompact,
	caption={Simulation code for Table 5.1},
	label={lst:table51}]
	# ============================================================
	# Simulation for Table 5.1
	# ============================================================
	rm(list = ls(all = TRUE))
	library(nleqslv)
	
	set.seed(123456)
	
	# ============================================================
	# Simulation settings
	# ============================================================
	a <- 3
	b <- 2
	n_ij <- 5
	alpha <- 0.01
	
	Q <- M <- 5000
	
	# Row and column shape parameters
	delta_i <- c(1, 0.5, 2)
	delta_j <- c(1, 2)
	
	# True shape parameters: delta_ij = delta_i * delta_j
	delta_in <- outer(delta_i, delta_j)
	beta_in <- 1
	
	# ============================================================
	# Counters and containers
	# ============================================================
	count <- 0                     # First call within each q
	count2 <- 0                    # PBLRT rejections
	count3 <- 0                    # ALRT rejections
	
	x <- array(NA_real_, c(a, b, n_ij))
	
	# Restricted estimates used for bootstrap generation
	delta_gen <- numeric(a + b)
	beta_gen <- NA_real_
	
	# ============================================================
	# Root-bracketing function
	# ============================================================
	bracket_root <- function(f, lower = .Machine$double.eps,
	upper = 5e4, max_expand = 8) {
		f_low <- suppressWarnings(tryCatch(f(lower), error = function(e) NA_real_))
		f_up  <- suppressWarnings(tryCatch(f(upper), error = function(e) NA_real_))
		if (!is.na(f_low) && !is.na(f_up) && f_low * f_up <= 0) return(c(lower, upper))
		
		# Expand the upper bound
		u <- upper
		for (k in 1:max_expand) {
			u <- u * 10
			f_up <- suppressWarnings(tryCatch(f(u), error = function(e) NA_real_))
			if (!is.na(f_low) && !is.na(f_up) && f_low * f_up <= 0) return(c(lower, u))
		}
		
		# Reduce the lower bound
		l <- lower
		for (k in 1:max_expand) {
			l <- max(l / 10, .Machine$double.eps)
			f_low <- suppressWarnings(tryCatch(f(l), error = function(e) NA_real_))
			if (!is.na(f_low) && !is.na(f_up) && f_low * f_up <= 0) return(c(l, u))
		}
		return(NA)
	}
	
	# ============================================================
	# Function to calculate the likelihood-ratio statistic
	# ============================================================
	cal_lambda <- function(a, b, n_ij, x) {
		
		# ----------------------------------------------------------
		# Arithmetic and geometric means
		# ----------------------------------------------------------
		x_bar <- matrix(NA_real_, a, b)
		x_tilde <- matrix(NA_real_, a, b)
		for (i in 1:a) for (j in 1:b) {
			x_bar[i, j] <- mean(x[i, j, ])
			x_tilde[i, j] <- exp(mean(log(x[i, j, ])))
		}
		
		# ==========================================================
		# 1. Unrestricted MLEs: Equations (3.6) and (3.7)
		# ==========================================================
		solve_unrestricted <- function(x_bar, x_tilde) {
			equations <- function(delta_vec) {
				delta_mat <- matrix(delta_vec, nrow = a, ncol = b)
				s1 <- sum(n_ij * delta_mat)
				s2 <- sum(n_ij * x_bar)
				
				eqs <- matrix(NA_real_, nrow = a, ncol = b)
				for (i in 1:a) for (j in 1:b) {
					eqs[i, j] <- log(s1) -
					digamma(delta_mat[i, j]) -
					log(s2 / x_tilde[i, j])
				}
				as.vector(eqs)
			}
			
			initial_guess <- rep(runif(1, 0.01, 1), a * b)
			
			sol <- try(
			nleqslv::nleqslv(initial_guess, equations, method = "Broyden",
			control = list(maxit = 2000, trace = 0)),
			silent = TRUE
			)
			
			if (inherits(sol, "try-error") || is.null(sol$termcd) ||
			sol$termcd != 1 || any(!is.finite(sol$x)) || any(sol$x <= 0))
			return(NA)
			
			matrix(sol$x, nrow = a, ncol = b)
		}
		
		delta <- solve_unrestricted(x_bar, x_tilde)
		if (any(is.na(delta))) return(NA_real_)
		
		beta <- sum(n_ij * x_bar) / sum(n_ij * delta)
		
		# ==========================================================
		# 2. Unrestricted log-likelihood: Equation (3.4)
		# ==========================================================
		L1 <- 0
		for (i in 1:a) for (j in 1:b) {
			L1 <- L1 + n_ij * (
			-lgamma(delta[i, j]) -
			delta[i, j] * log(beta) -
			x_bar[i, j] / beta +
			(delta[i, j] - 1) * log(x_tilde[i, j])
			)
		}
		
		# ==========================================================
		# 3. Restricted MLEs: Equations (5.2)--(5.4)
		# ==========================================================
		solve_restricted <- function(x_bar, x_tilde) {
			equations <- function(delta_vec) {
				eqs <- numeric(a + b)
				
				# Sum of restricted shape parameters
				t1 <- 0
				for (i in 1:a) for (j in 1:b) {
					t1 <- t1 + n_ij * delta_vec[i] * delta_vec[a + j]
				}
				t2 <- sum(n_ij * x_bar)
				
				# Equations for delta_i.
				for (i in 1:a) {
					eqs[i] <- sum(
					n_ij * delta_vec[(a + 1):(a + b)] *
					digamma(delta_vec[i] * delta_vec[(a + 1):(a + b)]) -
					n_ij * delta_vec[(a + 1):(a + b)] *
					log(x_tilde[i, ] * t1 / t2)
					)
				}
				
				# Equations for delta_.j
				for (j in 1:b) {
					eqs[a + j] <- sum(
					n_ij * delta_vec[1:a] *
					digamma(delta_vec[1:a] * delta_vec[a + j]) -
					n_ij * delta_vec[1:a] *
					log(x_tilde[, j] * t1 / t2)
					)
				}
				
				eqs
			}
			
			initial_guess <- runif(a + b, min = 0.1, max = 2)
			
			sol <- try(
			nleqslv::nleqslv(initial_guess, equations, method = "Broyden",
			control = list(maxit = 2000, trace = 0)),
			silent = TRUE
			)
			
			if (inherits(sol, "try-error") || is.null(sol$termcd) ||
			sol$termcd != 1 || any(!is.finite(sol$x)) || any(sol$x <= 0))
			return(NA)
			
			sol$x
		}
		
		delta_0 <- solve_restricted(x_bar, x_tilde)
		if (any(is.na(delta_0))) return(NA_real_)
		
		# Common scale parameter: Equation (5.4)
		delta_sum <- 0
		for (i in 1:a) for (j in 1:b) {
			delta_sum <- delta_sum + n_ij * delta_0[i] * delta_0[a + j]
		}
		beta_0 <- sum(n_ij * x_bar) / delta_sum
		
		# Store estimates from the observed sample
		if (count == 0) {
			delta_gen <<- delta_0
			beta_gen <<- beta_0
			count <<- 1
		}
		
		# ==========================================================
		# 4. Restricted log-likelihood: Equation (5.5)
		# ==========================================================
		L2 <- 0
		for (i in 1:a) for (j in 1:b) {
			delta_ij <- delta_0[i] * delta_0[a + j]
			
			L2 <- L2 + n_ij * (
			-lgamma(delta_ij) -
			delta_ij * log(beta_0) -
			x_bar[i, j] / beta_0 +
			(delta_ij - 1) * log(x_tilde[i, j])
			)
		}
		
		# ==========================================================
		# 5. Likelihood-ratio statistic
		# ==========================================================
		lambda_star <- 2 * (L1 - L2)
		return(lambda_star)
	}
	
	# ============================================================
	# Monte Carlo simulation
	# ============================================================
	start.time <- Sys.time()
	
	for (q in 1:Q) {
		
		# Reset the first-call indicator
		count <- 0
		
		# ==========================================================
		# Step 1: Generate the observed sample
		# ==========================================================
		repeat {
			for (i in 1:a) for (j in 1:b) {
				x[i, j, ] <- rgamma(n_ij, shape = delta_in[i, j], scale = beta_in)
			}
			
			lambda_star <- cal_lambda(a, b, n_ij, x)
			
			if (!is.na(lambda_star) && !is.infinite(lambda_star)) break
		}
		
		# ==========================================================
		# Step 2: Parametric bootstrap
		# ==========================================================
		lambda_rep <- numeric(M)
		
		for (m in 1:M) {
			repeat {
				xstar <- array(NA_real_, c(a, b, n_ij))
				
				for (i in 1:a) for (j in 1:b) {
					delta_star <- delta_gen[i] * delta_gen[a + j]
					
					xstar[i, j, ] <- rgamma(n_ij, shape = delta_star, scale = beta_gen)
				}
				
				resultm <- cal_lambda(a, b, n_ij, xstar)
				
				if (!is.na(resultm) && !is.infinite(resultm)) break
			}
			
			lambda_rep[m] <- resultm
		}
		
		# ==========================================================
		# Step 3: Bootstrap critical value
		# ==========================================================
		lambda_upper <- as.numeric(
		quantile(lambda_rep, probs = 1 - alpha, type = 7, na.rm = TRUE)
		)
		
		# ==========================================================
		# Step 4: ALRT and PBLRT
		# ==========================================================
		# Degrees of freedom used in the original simulation code
		df_LRT <- (a - 1) * (b - 1) - 1
		
		# Asymptotic likelihood-ratio test
		if (lambda_star > qchisq(1 - alpha, df = df_LRT)) count3 <- count3 + 1
		
		# Parametric bootstrap likelihood-ratio test
		if (lambda_star > lambda_upper) count2 <- count2 + 1
	}
	
	# ============================================================
	# Results
	# ============================================================
	p_ALRT <- count3 / Q
	p_PBLRT <- count2 / Q
	
	cat(
	"Q =", Q, "M =", M, "a =", a, "b =", b,
	"alpha =", alpha, "n_ij =", n_ij,
	"delta_i. =", delta_i, "delta_.j =", delta_j, "\n"
	)
	
	cat("ALRT count =", count3, "\n")
	cat("PBLRT count =", count2, "\n")
	cat("p_ALRT =", p_ALRT, "\n")
	cat("p_PBLRT =", p_PBLRT, "\n")
	
	end.time <- Sys.time()
	time.taken <- end.time - start.time
	print(time.taken)
	# ============================================================
	# End of script
	# ============================================================
\end{lstlisting}
\begin{lstlisting}[style=Rcompact,
	caption={Simulation code for Table 5.2},
	label={lst:table52}]
	# ============================================================
	# Simulation for Table 5.2
	# ============================================================
	rm(list = ls(all.names = TRUE))
	set.seed(123456)
	
	# ---------------------- Simulation settings ----------------------
	a     <- 3
	b     <- 2
	n_ij  <- 5
	alpha <- 0.01
	
	Q <- 5000                    # Number of Monte Carlo replications
	M <- 5000                    # Number of bootstrap replications
	
	# True Gamma shape parameters: delta_ij = delta_i * delta_j
	delta_i <- c(4, 5, 6)
	delta_j <- c(3, 4)
	delta_in <- outer(delta_i, delta_j)
	beta_in <- 1                 # True scale parameter
	
	# Rejection counters
	count_bootstrap <- 0          # PBLRT rejections
	count_asymptotic <- 0         # ALRT rejections
	
	# Bootstrap-generating parameter estimates
	mu_i_gen <- numeric(a)
	mu_j_gen <- numeric(b)
	beta_gen <- matrix(NA_real_, a, b)
	
	# =================================================================
	# Helper function: find an interval containing a root
	# =================================================================
	bracket_root <- function(f, lower = .Machine$double.eps,
	upper = 5e4, max_expand = 8) {
		f_low <- suppressWarnings(tryCatch(f(lower), error = function(e) NA_real_))
		f_up  <- suppressWarnings(tryCatch(f(upper), error = function(e) NA_real_))
		if (is.finite(f_low) && is.finite(f_up) && f_low * f_up <= 0)
		return(c(lower, upper))
		
		# Expand the upper bound
		u <- upper
		for (k in seq_len(max_expand)) {
			u <- 10 * u
			f_up <- suppressWarnings(tryCatch(f(u), error = function(e) NA_real_))
			if (is.finite(f_low) && is.finite(f_up) && f_low * f_up <= 0)
			return(c(lower, u))
		}
		
		# Reduce the lower bound
		l <- lower
		for (k in seq_len(max_expand)) {
			l <- max(l / 10, .Machine$double.eps)
			f_low <- suppressWarnings(tryCatch(f(l), error = function(e) NA_real_))
			if (is.finite(f_low) && is.finite(f_up) && f_low * f_up <= 0)
			return(c(l, u))
		}
		NA_real_
	}
	
	# =================================================================
	# Compute arithmetic and geometric means for each cell
	# =================================================================
	cal_AM_GM <- function(a, b, x) {
		x_bar <- matrix(NA_real_, a, b)
		x_tilde <- matrix(NA_real_, a, b)
		for (i in seq_len(a)) for (j in seq_len(b)) {
			xx <- x[i, j, ]
			x_bar[i, j] <- mean(xx)
			x_tilde[i, j] <- exp(mean(log(xx)))
		}
		list(x_bar = x_bar, x_tilde = x_tilde)
	}
	
	# =================================================================
	# Estimate unrestricted parameters and log-likelihood
	# =================================================================
	estimate_unrestricted <- function(a, b, n_ij, x_bar, x_tilde) {
		delta <- matrix(NA_real_, a, b)
		beta  <- matrix(NA_real_, a, b)
		
		for (i in seq_len(a)) for (j in seq_len(b)) {
			equation <- function(d) digamma(d) - log(d) - log(x_tilde[i, j] / x_bar[i, j])
			
			interval <- bracket_root(equation)
			if (anyNA(interval)) return(NULL)
			
			root <- suppressWarnings(
			tryCatch(uniroot(equation, interval = interval)$root, error = function(e) NA_real_)
			)
			if (!is.finite(root) || root <= 0) return(NULL)
			
			delta[i, j] <- root
			beta[i, j]  <- x_bar[i, j] / root
		}
		
		loglik <- sum(
		n_ij * (
		-lgamma(delta) -
		delta * log(beta) -
		x_bar / beta +
		(delta - 1) * log(x_tilde)
		)
		)
		
		list(delta = delta, beta = beta, loglik = loglik)
	}
	
	# =================================================================
	# Pack and unpack restricted-model parameters
	# =================================================================
	pack <- function(mu_i, mu_j, beta) c(mu_i, mu_j, as.vector(beta))
	
	unpack <- function(theta, a, b) {
		mu_i <- theta[seq_len(a)]
		mu_j <- theta[a + seq_len(b)]
		
		beta_start <- a + b + 1
		beta <- matrix(theta[beta_start:(beta_start + a * b - 1)], nrow = a, ncol = b)
		
		list(mu_i = mu_i, mu_j = mu_j, beta = beta)
	}
	
	# =================================================================
	# Restricted likelihood equations: Equations (5.9)--(5.11)
	# =================================================================
	equations <- function(theta, n_ij, a, b, x_bar, x_tilde) {
		parameters <- unpack(theta, a, b)
		mu_i <- parameters$mu_i
		mu_j <- parameters$mu_j
		beta <- parameters$beta
		
		eq <- numeric(a + b + a * b)
		
		# Equation (5.9)
		for (i in seq_len(a)) {
			eq[i] <- sum(
			n_ij * mu_j / beta[i, ] *
			(digamma(mu_i[i] * mu_j / beta[i, ]) - log(x_tilde[i, ] / beta[i, ]))
			)
		}
		
		# Equation (5.10)
		for (j in seq_len(b)) {
			eq[a + j] <- sum(
			n_ij * mu_i / beta[, j] *
			(digamma(mu_i * mu_j[j] / beta[, j]) - log(x_tilde[, j] / beta[, j]))
			)
		}
		
		# Equation (5.11)
		k <- a + b
		for (i in seq_len(a)) for (j in seq_len(b)) {
			k <- k + 1
			shape <- mu_i[i] * mu_j[j] / beta[i, j]
			
			eq[k] <- digamma(shape) + log(beta[i, j] / x_tilde[i, j]) -
			1 + x_bar[i, j] / (mu_i[i] * mu_j[j])
		}
		
		eq
	}
	
	# =================================================================
	# Estimate restricted parameters under H0
	# =================================================================
	estimate_restricted <- function(a, b, n_ij, x_bar, x_tilde, max_tries = 30) {
		for (attempt in seq_len(max_tries)) {
			
			# Use deterministic starting values first and random
			# starting values in subsequent attempts
			if (attempt == 1) {
				mu_i0 <- sqrt(rowMeans(x_bar))
				mu_j0 <- sqrt(colMeans(x_bar))
				beta0 <- matrix(1, a, b)
			} else {
				mu_i0 <- runif(a, 0.1, 3)
				mu_j0 <- runif(b, 0.1, 3)
				beta0 <- matrix(runif(a * b, 0.1, 3), a, b)
			}
			
			theta0 <- pack(mu_i0, mu_j0, beta0)
			
			fit <- try(
			nleqslv::nleqslv(
			x = theta0, fn = equations, n_ij = n_ij, a = a, b = b,
			x_bar = x_bar, x_tilde = x_tilde,
			method = "Broyden", control = list(maxit = 2000)
			),
			silent = TRUE
			)
			
			if (inherits(fit, "try-error") || fit$termcd != 1) next
			
			solution <- unpack(fit$x, a, b)
			
			if (any(solution$mu_i <= 0) || any(solution$mu_j <= 0) || any(solution$beta <= 0))
			next
			
			mu_i <- solution$mu_i
			mu_j <- solution$mu_j
			beta <- solution$beta
			
			shape <- outer(mu_i, mu_j) / beta
			
			loglik <- sum(
			n_ij * (
			-lgamma(shape) -
			shape * log(beta) -
			x_bar / beta +
			(shape - 1) * log(x_tilde)
			)
			)
			
			return(list(mu_i = mu_i, mu_j = mu_j, beta = beta, loglik = loglik))
		}
		
		NULL
	}
	
	# =================================================================
	# Compute the likelihood-ratio statistic Lambda*
	# =================================================================
	cal_lambda <- function(a, b, n_ij, x_bar, x_tilde) {
		unrestricted <- estimate_unrestricted(a, b, n_ij, x_bar, x_tilde)
		if (is.null(unrestricted)) return(NULL)
		
		restricted <- estimate_restricted(a, b, n_ij, x_bar, x_tilde)
		if (is.null(restricted)) return(NULL)
		
		if (!is.finite(unrestricted$loglik) || !is.finite(restricted$loglik))
		return(NULL)
		
		list(
		lambda_star = -2 * (restricted$loglik - unrestricted$loglik),
		mu_i = restricted$mu_i,
		mu_j = restricted$mu_j,
		beta = restricted$beta
		)
	}
	
	# =================================================================
	# Main simulation
	# =================================================================
	start.time <- Sys.time()
	
	for (q in seq_len(Q)) {
		
		# ---------------------------------------------------------------
		# Step 1: Generate the original sample and compute Lambda*
		# ---------------------------------------------------------------
		repeat {
			x <- array(NA_real_, c(a, b, n_ij))
			
			for (i in seq_len(a)) for (j in seq_len(b)) {
				x[i, j, ] <- rgamma(n_ij, shape = delta_in[i, j], scale = beta_in)
			}
			
			means <- cal_AM_GM(a, b, x)
			result <- cal_lambda(a, b, n_ij, means$x_bar, means$x_tilde)
			
			if (!is.null(result) && is.finite(result$lambda_star)) break
		}
		
		lambda_star <- result$lambda_star
		
		# Save restricted estimates for parametric bootstrap
		mu_i_gen <- result$mu_i
		mu_j_gen <- result$mu_j
		beta_gen <- result$beta
		
		# ---------------------------------------------------------------
		# Step 2: Parametric bootstrap
		# ---------------------------------------------------------------
		lambda_rep <- numeric(M)
		
		for (m in seq_len(M)) {
			repeat {
				xstar <- array(NA_real_, c(a, b, n_ij))
				
				for (i in seq_len(a)) for (j in seq_len(b)) {
					shape_gen <- mu_i_gen[i] * mu_j_gen[j] / beta_gen[i, j]
					
					xstar[i, j, ] <- rgamma(n_ij, shape = shape_gen, scale = beta_gen[i, j])
				}
				
				means_star <- cal_AM_GM(a, b, xstar)
				result_star <- cal_lambda(a, b, n_ij, means_star$x_bar, means_star$x_tilde)
				
				if (!is.null(result_star) && is.finite(result_star$lambda_star)) break
			}
			
			lambda_rep[m] <- result_star$lambda_star
		}
		
		# ---------------------------------------------------------------
		# Step 3: ALRT and PBLRT decisions
		# ---------------------------------------------------------------
		lambda_upper <- as.numeric(
		quantile(lambda_rep, probs = 1 - alpha, type = 7)
		)
		
		df <- (a - 1) * (b - 1) - 1
		
		if (lambda_star > qchisq(1 - alpha, df = df)) count_asymptotic <- count_asymptotic + 1
		if (lambda_star > lambda_upper) count_bootstrap <- count_bootstrap + 1
	}
	
	# =================================================================
	# Simulation results
	# =================================================================
	p_ALRT  <- count_asymptotic / Q
	p_PBLRT <- count_bootstrap / Q
	
	cat(
	"Q =", Q, "M =", M, "a =", a, "b =", b,
	"alpha =", alpha, "n_ij =", n_ij, "\n"
	)
	
	cat("delta_i =", delta_i, "\n")
	cat("delta_j =", delta_j, "\n")
	cat("ALRT rejections =", count_asymptotic, "\n")
	cat("PBLRT rejections =", count_bootstrap, "\n")
	cat("p_ALRT =", p_ALRT, "\n")
	cat("p_PBLRT =", p_PBLRT, "\n")
	
	end.time <- Sys.time()
	time.taken <- end.time - start.time
	print(time.taken)
	# ======================== End of Script ========================
\end{lstlisting}
\section{The $p$-value computations using the gamma model for Example 1.1}
\begin{lstlisting}[style=Rcompact,
	caption={Problem - 5 for the Example 1.1},
	label={lst:table22data1}]
	# ============================================================
	# Table 2.2: Example 1.1
	# ============================================================
	rm(list = ls(all = TRUE))
	library(nleqslv)
	
	a <- 2
	b <- 6
	n_ij <- 5
	alpha <- 0.05
	M <- 5000
	
	count  <- 0  # Save restricted estimates from the original data
	count2 <- 0  # Number of bootstrap statistics exceeding lambda_star
	
	Gender <- factor(rep(c("1", "2"), each = 6 * 5))
	Hormone <- factor(rep(rep(1:6, each = 5), times = 2))
	value <- c(
	# Gender 1
	274.99, 289.67, 346.40, 344.32, 364.63,
	478.62, 399.14, 512.03, 400.74, 369.48,
	276.32, 318.32, 310.48, 352.29, 377.58,
	391.45, 351.85, 299.05, 288.55, 259.10,
	405.58, 318.63, 291.58, 311.40, 297.80,
	264.89, 308.36, 338.21, 352.18, 376.36,
	# Gender 2
	153.96, 153.92, 181.88, 132.75, 117.50,
	190.02, 253.14, 176.91, 227.31, 237.62,
	111.37, 139.92, 166.74, 140.66, 141.31,
	120.05, 127.68, 145.49, 172.93, 193.85,
	118.60, 153.60, 139.52, 159.76, 172.52,
	149.38, 143.54, 124.30, 164.93, 177.85
	)
	
	data <- data.frame(Gender, Hormone, value)
	x <- aperm(array(data$value, c(n_ij, b, a)), c(3, 2, 1))
	
	delta_gen <- matrix(NA_real_, a, b)
	beta_gen <- numeric(1)
	
	# Bracket a root for uniroot()
	bracket_root <- function(f, lower = .Machine$double.eps,
	upper = 5e4, max_expand = 8) {
		f_low <- suppressWarnings(tryCatch(f(lower), error = function(e) NA_real_))
		f_up  <- suppressWarnings(tryCatch(f(upper), error = function(e) NA_real_))
		if (!is.na(f_low) && !is.na(f_up) && f_low * f_up <= 0) return(c(lower, upper))
		
		u <- upper
		for (k in 1:max_expand) {
			u <- u * 10
			f_up <- suppressWarnings(tryCatch(f(u), error = function(e) NA_real_))
			if (!is.na(f_low) && !is.na(f_up) && f_low * f_up <= 0) return(c(lower, u))
		}
		
		l <- lower
		for (k in 1:max_expand) {
			l <- max(l / 10, .Machine$double.eps)
			f_low <- suppressWarnings(tryCatch(f(l), error = function(e) NA_real_))
			if (!is.na(f_low) && !is.na(f_up) && f_low * f_up <= 0) return(c(l, u))
		}
		NA
	}
	
	# Calculate the likelihood-ratio statistic
	cal_lambda <- function(a, b, n_ij, x) {
		x_bar <- matrix(NA_real_, a, b)
		x_tilde <- matrix(NA_real_, a, b)
		for (i in 1:a) for (j in 1:b) {
			xx <- x[i, j, ]
			x_bar[i, j] <- mean(xx, na.rm = TRUE)
			x_tilde[i, j] <- exp(mean(log(xx), na.rm = TRUE))
		}
		
		delta <- matrix(NA_real_, a, b)
		beta <- matrix(NA_real_, a, b)
		for (i in 1:a) for (j in 1:b) {
			f1 <- function(z) digamma(z) - log(z) - log(x_tilde[i, j] / x_bar[i, j])
			
			br <- bracket_root(f1)
			if (all(is.na(br))) return(NA_real_)
			
			root <- suppressWarnings(
			tryCatch(uniroot(f1, interval = br)$root, error = function(e) NA_real_)
			)
			if (is.na(root)) return(NA_real_)
			
			delta[i, j] <- root
			beta[i, j] <- x_bar[i, j] / root
		}
		
		L1 <- 0
		for (i in 1:a) for (j in 1:b) {
			L1 <- L1 + n_ij * (
			-lgamma(delta[i, j]) -
			delta[i, j] * log(beta[i, j]) -
			x_bar[i, j] / beta[i, j] +
			(delta[i, j] - 1) * log(x_tilde[i, j])
			)
		}
		
		equations <- function(delta_vec) {
			delta_mat <- matrix(delta_vec, a, b)
			eqs <- matrix(NA_real_, a, b)
			for (i in 1:a) for (j in 1:b)
			eqs[i, j] <- digamma(delta_mat[i, j]) -
			log(sum(n_ij * delta_mat)) -
			log(x_tilde[i, j] / sum(n_ij * x_bar))
			as.vector(eqs)
		}
		
		initial_num <- runif(1, 0.001, 1)
		initial_guess <- rep(initial_num, a * b)
		
		sol <- try(
		nleqslv(initial_guess, equations, method = "Broyden",
		control = list(maxit = 2000, trace = 0)),
		silent = TRUE
		)
		if (inherits(sol, "try-error") || is.null(sol$termcd) || sol$termcd != 1)
		return(NA_real_)
		
		delta_0 <- matrix(sol$x, a, b)
		beta_0 <- sum(n_ij * x_bar) / sum(n_ij * delta_0)
		
		if (count == 0) {
			delta_gen <<- delta_0
			beta_gen <<- beta_0
			count <<- count + 1
		}
		
		L0 <- 0
		for (i in 1:a) for (j in 1:b) {
			L0 <- L0 + n_ij * (
			-lgamma(delta_0[i, j]) -
			delta_0[i, j] * log(beta_0) -
			x_bar[i, j] / beta_0 +
			(delta_0[i, j] - 1) * log(x_tilde[i, j])
			)
		}
		
		lambda <- exp(L0 - L1)
		-2 * log(lambda)
	}
	
	# Compute the observed likelihood-ratio statistic
	start.time <- Sys.time()
	count <- 0
	lambda_star <- cal_lambda(a, b, n_ij, x)
	
	# Parametric bootstrap
	lambda_rep <- rep(NA_real_, M)
	for (m in 1:M) {
		repeat {
			xstar <- array(NA_real_, c(a, b, n_ij))
			for (i in 1:a) for (j in 1:b) for (k in 1:n_ij)
			xstar[i, j, k] <- rgamma(1, shape = delta_gen[i, j], scale = beta_gen)
			
			resultm <- cal_lambda(a, b, n_ij, xstar)
			if (!is.na(resultm) && !is.infinite(resultm)) break
		}
		
		lambda_rep[m] <- resultm
		if (lambda_rep[m] > lambda_star) count2 <- count2 + 1
	}
	
	cat("Dataset 1:", "M =", M, "a =", a, "b =", b,
	"alpha =", alpha, "n_ij =", n_ij, "\n")
	cat("lam_star =", lambda_star, "\n")
	
	p_ALRT <- 1 - pchisq(lambda_star, a * b - 1)
	p_PBLRT <- count2 / M
	
	cat("p_ALRT =", p_ALRT, "\n")
	cat("p_PBLRT =", p_PBLRT, "\n")
	
	end.time <- Sys.time()
	time.taken <- end.time - start.time
	print(time.taken)
	sink()
	# ====================== End of Script ======================
\end{lstlisting}

}

\end{document}